\documentclass[a4paper,11pt]{article}
\usepackage{jcappub} 

\pdfoutput=1

\usepackage{graphicx}	
\usepackage{amsmath}	
\usepackage{aas_macros}
\usepackage{subcaption}
\usepackage{dblfloatfix} 
\usepackage{float}
\usepackage{aas_macros}

\def\be{\begin{equation}}
\def\ee{\end{equation}}
\def\beq{\begin{eqnarray}}
\def\eeq{\end{eqnarray}}
\def\mr{\mathrm}

\title{Simulating the epoch of Helium Reionization in photon-conserving semi-numerical code SCRIPT}

\author[a,1]{Akanksha Kapahtia \note{Corresponding author.}}
\author[a]{and T. Roy Choudhury}

\affiliation[a]{National Centre for Radio Astrophysics (TIFR), \\
Pune University Campus, Ganeshkhind, Pune 411007, India}

\emailAdd{akanksha.kapahtia@gmail.com}
\emailAdd{tirth@ncra.tifr.res.in}

\abstract{The reionization of the second electron of helium (HeII) leaves important imprints on the thermal and ionization state of the intergalactic medium (IGM). Observational evidence suggests that HeII reionization ended at $z \simeq 3$ due to ionizing photons emitted predominantly by quasars. We present efficient semi-numerical simulations of helium reionization in a $230 \ \mathrm{h^{-1}~Mpc}$ box, that takes into account the spatial patchiness of reionization coupled with photoheating of the IGM. Dark matter haloes are assigned quasars using empirical measurements of the quasar luminosity function, assuming a universal quasar lifetime consistent with duty cycle values inferred from measurements of the quasar clustering. The ionizing photon field from quasars is then included in the semi-numerical Code for ReionIzation with PhoTon conservation (SCRIPT), which was originally developed for modeling hydrogen reionization. In this work, we make appropriate modifications to SCRIPT for modeling inhomogenous HeII reionization and the corresponding thermal history of the IGM is modelled via a subgrid prescription. Our model has three main free parameters i.e. the global clumping factor $\mathcal{C}_{HeIII}$, the temperature increase due to photoheating $T^{re}_{He}$ and the quasar spectral energy distribution (SED) index, $\alpha_{UV}$. Our \textit{fiducial} model with $\mathcal{C}_{HeIII}=15.6$ and  $T^{re}_{He} \sim 6000 \ K$ gives reasonable values for the empirical measurements of the temperature density equation of state at these redshifts, assuming that quasars brighter than $\mr{M_{1450}}<-21$ and having $\alpha_{UV}=1.7$ contribute to HeII reionization. The efficiency of our code shows promising prospects for performing parameter estimation in future, for models of HeII reionization using observations of the Ly$\alpha$ forest.  }

\begin{document}
\maketitle
\flushbottom
\section{Introduction}

The evolution of the intergalactic medium (IGM) in the post cosmic dawn era is characterized by two major baryonic phase transitions. The first of these transitions corresponds to the ionization of intergalactic hydrogen (and the first electron of helium) \citep{Dayal2018, Gnedin,Choudhury2022}, while the second transition occurs at a later cosmic time due to the ionization of singly ionized helium. The relatively high ionization energy of HeII ($\mathrm{E_{\gamma}=54.4 \ eV}$) and faster recombinations ($\sim 5.5$ times faster than the recombination of HII), requires ionizing photons from sources with a harder spectra. Since a significant fraction of such photons is produced by quasars \citep{Sanderbeck_2018}, helium reionization occurs at a later epoch ($2.5  \lesssim z \lesssim 4 $) when the quasar number densities become high enough to produce the bulk of HeII ionizing photons \citep{Wyithe_2003, Girish_UV}. 

 Observationally, the most direct probe of helium reionization is the HeII Ly$\alpha$ forest, seen bluewards of the HeII Ly$\alpha$ resonance at $304$ \AA ~ in the spectra of distant ($z \gtrsim 2$) quasars. A small fraction of HeII is sufficient to produce a HeII Gunn-Peterson trough \citep{Miralda,fardal1998high}. Therefore, the HeII forest allows for probing the tail end of helium reionization. The trough was first observed in the far UV along the line of sight of quasar Q0302-003 at z=3.285 \citep{Jakobsen1994,madau1994he} by the Hubble Space Telescope's (HST) \textit{Faint Objective Camera  (FOC)}.  Quantitative measurements of the HeII forest along different quasar sightlines were later enabled by the HST's \textit{Cosmic Origin's Spectrograph (COS)} and the \textit{Far Ultraviolet Spectroscopic Explorer(FUSE)} satellite. The HeII opacity along multiple sightlines is statistically measured in terms of the effective optical depth $\tau_{eff}$. The number of quasars that show emission at HeII Ly$\alpha$ resonance is limited by the decreasing quasar number densities with increasing redshifts ($z \gtrsim 3$) and intervening systems with Lyman continuum absorption (at $ \lambda \ge 912$ \AA). As a result, so far there are only $25$ high signal to noise HeII sightlines at $2.3 \lesssim z \lesssim 3.8$ which have been used for statistical study of the HeII Ly$\alpha$ forest probing the end stages of helium reionization \citep{Davidsen1996, Heap_2000,Smette_2002, Zheng2008, shull2010hst, Worseck11,Worseck2016}. The sightline to sightline variation of $\tau_{eff}$ for this sample indicates that HeII reionization is an extended process that ended at $z \lesssim  2.7$ \citep{Worseck19}. This was further corroborated by recent measurements of the transmission spikes of the HeII Ly$\alpha$ forest \citep{Makan_2021,Makan_2022} for about 10 of these sightlines between $2.5 \lesssim z \lesssim 3.8$, which indicated that HeII reionization had progressed sufficiently at these redshifts.
 
Given that quasars are the main drivers of helium reionization, their relatively lower prevalence along with the inhomogeneity of the IGM makes helium reionization a patchy process. The metagalactic UV background from galaxies and quasars is responsible for keeping the IGM ionized after reionization. The UV background models are an important input to hydrodynamical simulations of structure formation to efficiently model the ionization and thermal state of the IGM \citep{Onorbe2017,Puchwein2019,Faucher-Giguere20,villasenor2022inferring}. The patchy nature of helium reionization induces large fluctuations in this UV background. These fluctuations in the ionizing UV background have important implications for interpreting the Ly$\alpha$ forest measurements at $z \gtrsim 2$ \citep{Davies2014,Davies2017, Meiksin20}. Additionally, the inhomogenous photoheating of the IGM due to photons having energies in excess of the ionization energy of HeII raises the temperature of the gas considerably and is a dominant heating mechanism especially at low densities relevant for Ly$\alpha$ forest studies. This heating impacts the width and depth of the small scale features in the HI Ly$\alpha$ forest. A similar heating event occurs during hydrogen reionization \citep[e.g.][]{Prakash2020}. After the epoch of hydrogen reionization, the temperature and density of the adiabatically expanding ionized gas at low densities ($\rho/\bar{\rho} \lesssim 10$) settles down to a tight power law relation \citep{miralda1994reionization,Hui1997,hui_haiman,Mcquinn2016,Sanderbeck_2018}.  Statistics of the HI Ly$\alpha$ forest are used for putting constraints on the temperature-density equation of state of the IGM, which would be altered due to the subsequent photoheating of the IGM during helium reionization \citep{Schaye2000,lidz2010measurement,Becker2011,boera2014thermal}. The most recent measurements of the amplitude and slope of the temperature density equation of state show a peak in the temperature evolution at $z \sim 3$ \citep{Prakash21} attributed to helium reionization. 

The impact of helium reionization on the structure of the IGM makes it pertinent to model the sources, timing and duration of this epoch in a way that it is consistent with observational constraints. A self consistent modelling of helium reionization must incorporate the impact of inhomogenous ionization and heating of the IGM across a wide range of scales. On one hand, this requires modelling scales as large as a Gpc in order to account for the clustering properties of quasars which determine the topology of helium reionization. On the other hand, such models must be able to reproduce features in the Ly$\alpha$ forest at scales as small as $ \sim 100 \ \mathrm{kpc}$, since it is the only direct observable. The wide variety of scales involved makes a detailed modeling computationally expensive. Hence, studies that have attempted numerical techniques to model helium reionization, mostly consisted of hydrodynamical or N-body simulations post-processed with radiative transfer \citep{Sokasian,Paschos2007,Mcquinn2009}, but either excluded calculations of the coupled gas temperature evolution or inhomogeneous recombination which have important consequence for the Ly$\alpha$ forest measurements. These limitations are overcome by using techniques such as the adaptive mesh refinement \cite{Compostella_2013}. Alternatively, there are studies which use a more realistic approach of coupled hydrodynamic and radiative transfer simulations in smaller ($25 \ \mathrm{h^{-1} \ Mpc}$) \citep{Meiksin_tittley} boxes. While the small box simulation realistically captures the hydrodynamical response of gas due to heating during helium reionzation at resolutions relevant for Ly$\alpha$ forest, it ignores the large scale topology of reionization. Conversely,  a similar approach in larger boxes ($200 \mathrm{h^{-1} \ Mpc}$ \citep{Plante1, Plante2}), captures realistic details of the patchiness of helium reionization but ignores the small scale gas physics. Thus, spanning the entire dynamic range in realistic simulations is not feasible. 

A fair insight into the large scale morphology of the process have been obtained in semi-analytical models \citep{Furlanetto,Furlanetto_2008} or Monte Carlo based semi-analytical techniques \citep{Gleser2005}, but these techniques lack in their ability to incorporate spatial information, particularly that arising due to clustering of quasars.  Another approach to model helium reionization is by  using spatially uniform ionizing background models \citep[e.g.][]{Puchwein2019,Faucher-Giguere20} over hydrodynamic simulations. Since these models ignore spatial patchiness in heating and ionization, a thermal physics model was introduced to mimic radiative transfer on hydrodynamical simulations \cite{Sanderbeck2020}. While the model introduces patchiness, the treatment of ionization and IGM heating evolution is not coupled and relies on reionization pre-computed in a separate radiative transfer code. Thus, due to their computational inefficiency or lack of self-consistent modeling, most numerical and analytical treatments of helium reionization do not allow for exploring the wide range of parameters spanning the IGM and uncertain source properties. One way to circumvent this issue is to use a semi-numerical approach. While efficient, a previously introduced semi-numerical model \citep{Dixon_2014} suffered from some drawbacks, mainly as it lacked tracking redshift evolution of the ionization fraction, photoionization heating and a treatment for recombinations which are particularly important between quasar episodes. 

In this paper, we introduce an efficient semi-numerical approach for modelling helium reionization which self-consistently models inhomogenous recombinations and coupled thermal fluctuations. For this purpose, we use \textbf{S}eminumerical \textbf{C}ode for \textbf{R}e\textbf{I}onization with \textbf{P}ho\textbf{T}on conservation (SCRIPT) \citep{Tirth_script}, which was originally developed for modeling hydrogen reionization and modify it for helium reionization. Due to the numerical convergence of SCRIPT with respect to resolution, the relevent results can be obtained by running the simulations on a coarse grid (instead of higher resolutions), which leads to an increase in computational efficiency. However, this also means that most of the small scale physics (which is not captured at low resolutions), needs to be modelled using sub-grid and (semi-)analytical prescriptions. Therefore, the thermal history in our box is modelled analytically in a sub-grid fashion. Our \textit{fiducial} model which is consistent with empirical measurements of the quasar properties, reproduces the average measurements of the IGM equation of state. The main advantage of our semi-numerical simulation is its computational efficiency while incorporating a reasonable description of the reionization physics, thus serving as a compromise between detailed numerical simulations and analytical/semi-analytical techniques. Furthermore, its speed shows promising prospects for parameter estimation using the available observations of helium reionization. 

The plan of the paper is as follows: In Section~\ref{sec.2} we describe our procedure for modelling quasars as sources of helium reionization, which are then incorporated in SCRIPT modified for helium reionization, described in Section~\ref{sec.3}. We describe the redshift evolution of ionization and subgrid implementation of our thermal histories in Section~\ref{sec.4}. Finally, in Section~\ref{sec.5} we discuss the prospects of extending our model for $\tau_{eff}$ measurements to perform parameter estimation in future and conclude. The cosmological parameters used throughout our calculations are those for a flat $\Lambda$CDM cosmology given by: $\Omega_m=0.308$, $\Omega_b=0.0482$, $h=0.678$, $n_s=0.96$ and $\sigma_8=0.81$ \citep{Planck16} and the fraction of helium by mass, $Y_P=0.24$. 

\section{Modelling quasars as sources} \label{sec.2}

The main challenge in modelling quasars as HeII reionization sources is to assign quasar properties to dark matter haloes that fit well with the observed constraints, some of which are uncertain. While global properties like the quasar luminosity function and clustering measurements are relatively well measured, properties like the quasar spectral index and quasar lifetimes are highly uncertain \citep{qsr_lt_dist}. In this work, we use a data-driven approach to model the quasars. We first generate a dark matter halo catalogue using an $N$-body simulation and then populate these haloes with quasars having properties which are consistent with observations within the limitations of our box size. For our halo catalogue, we use the publicly available code \texttt{GADGET-4} \citep{gadget4} with $512^3$ particles in a box of length $230~h^{-1} \mathrm{cMpc}$. We enable the on-the-fly FoF in \texttt{GADGET-4} with a minimum group length of 20 particles leading to a minimum halo mass of $2 \times 10^{11} \ \mr{M}_\odot$. The ionizing photons at a given redshift from all the quasars in the box would be determined by the number of active quasars, their individual ionizing luminosity and timescale. We describe how these properties are modelled, below:

\subsection{Assigning quasars to haloes}
\begin{figure}  
 \centering
  \includegraphics[height=8.cm,width=8.cm]{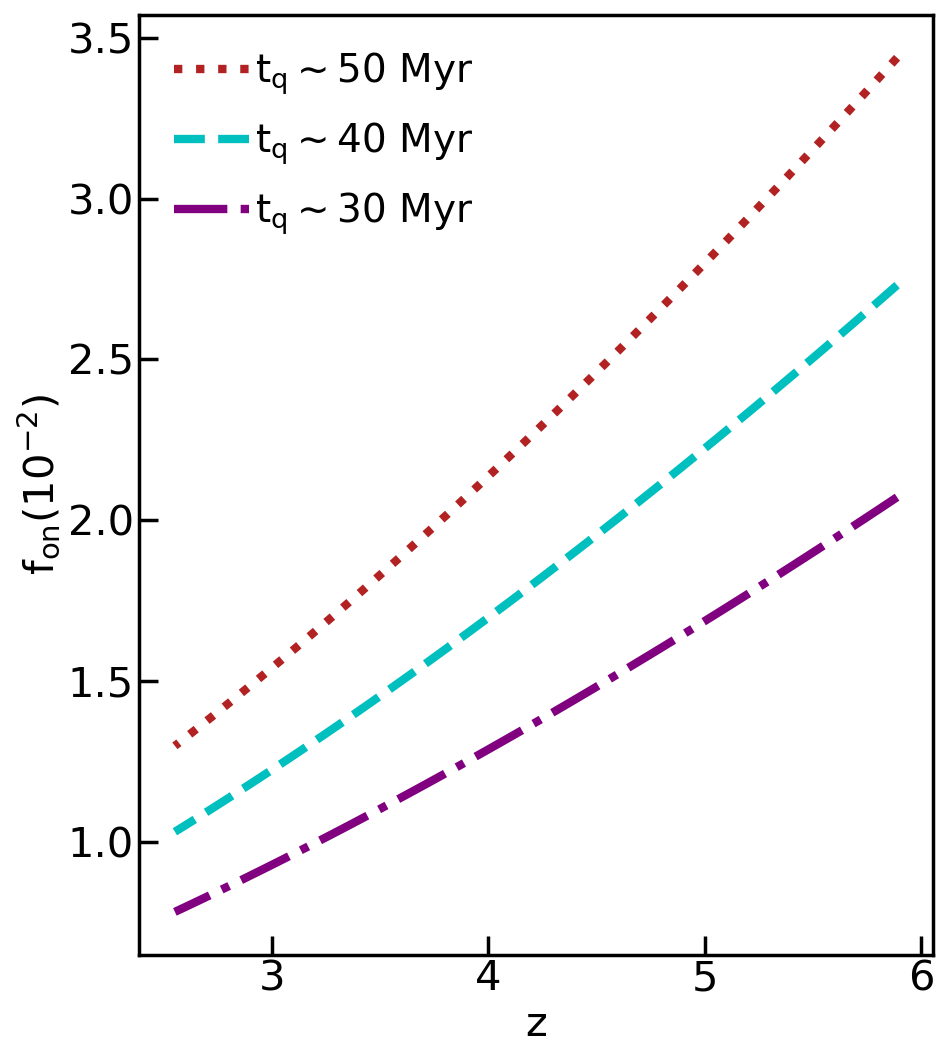}
    \caption{Evolution of the quasar duty cycle $f_\mathrm{ON}(z)$ for different values of quasar lifetimes $t_q$.}
    \label{tq_duty}
\end{figure}

At a given time all haloes capable of hosting a quasar cannot host an active quasar since they are active only for a short duration. Therefore, in order to assign quasars to dark matter haloes, we first need to specify the fraction of host haloes which are active at a certain redshift, or the quasar duty-cycle $f_\mathrm{ON}(z)$ which is defined as :
\beq
f_\mathrm{ON}(z) \equiv \cfrac{n_\mathrm{qso}}{\int_{M_\mathrm{min}}^{\infty} \cfrac{dn_\mathrm{{halo}} (M)}{dM} {dM}} = \cfrac{t_q}{t_H(z)}
\eeq
In the first equality of the above equation, $n_\mathrm{qso}$ is the number density of quasars and the denominator is the integral over the halo mass function, with $M_\mathrm{min}$ denoting the minimum mass for a halo capable of hosting a quasar. The second equality shows that $f_\mathrm{ON}$ is simply the ratio of the time over which a quasar is active, known as the quasar lifetime $t_q$, and the Hubble timescale $t_H(z) \equiv H^{-1}(z)$. In this work, we assume that $t_q$ and hence $f_\mathrm{ON}$ is independent of the halo mass. The evolution of $f_\mathrm{ON}(z)$ for different values of $t_q$ is shown in Figure~\ref{tq_duty}.

Given the halo catalogue and $f_\mathrm{ON}$, we assign luminosities using \textit{abundance matching}. As the name suggests, abundance matching relies on the \textit{ansatz} that the quasar luminosity is a monotonic function of the halo mass, due to which the cumulative number of quasars above a certain luminosity would be proportional to the cumulative number of haloes above a corresponding mass. 
Quasar clustering measurements at $2.5 \lesssim z \lesssim 4.0$ show that the quasar bias is almost independent of quasar luminosity \cite{Eftekharzadeh2015,He}. This probably indicates that there is some scatter between quasar luminosity and host halo mass, although it has been found to be moderate \citep[e.g.,][]{White_2012}. Unfortunately, measurements of luminosity-dependent clustering are rare at higher redshifts that are more relevant for our work. Therefore, we take the simplifying assumption of a monotonic relation between quasar luminosity and mass of the host halo.

The method can be expressed for an observed luminosity function $\Phi(L)$ and halo mass function $\cfrac{dn_\mathrm{halo}(M)}{dM}$ as:
\beq
\int_L^{\infty} dL'~\Phi(L') = f_\mathrm{ON}(z) \int_{M_{\mathrm{halo}}}^{\infty} d M'~\cfrac{dn_\mathrm{halo}(M')}{dM'}.
\label{ab}
\eeq
The above relation defines an implicit relation between $L$ and $\mr{M_{halo}}$, thus allowing us to assign luminosities to the haloes in the catalogue. 
The observed luminosity function is assumed to have a double power law parameterised by: (i) amplitude: $\phi^*$, (ii) break magnitude: $M^*_{1450}$, (iii) bright end slope: $\alpha$ and (iv) faint end slope: $\beta$. The evolution of the parameters is implemented using Model 2 of the luminosity function measurements given in \citep{Girish_UV} (see Appendix \ref{A1} for more details). The model has a total of 14 parameters and accounts for the break at $z \sim 3$ that occurs due to lack of credible data at these redshifts. The model also excludes data which has approximate selection functions. Since the empirical luminosity function is obtained for a quasar sample homogenized to $1450$ \AA ~ rest frame continuum, our abundance matching procedure would yield the absolute magnitude, $\mr{M_{1450}}$ (or $\mr{L_{1450}}=10^{-0.4\ [\mr{M_{1450}} -51.60]}$) for the corresponding halo mass $\mr{M_{halo}}$ from equation~\ref{ab}.

We choose $t_q$ to be in the range $30 - 50$~Myr. These values lead to large-scale clustering of quasars consistent with recent measurements \citep{White_2012,Eftekharzadeh2015}. We also choose to work with a simple \textit{lightbulb} model where the quasar luminosity remains constant over its lifetime. To construct the quasar catalogue, we choose a random number derived from a uniform distribution for each halo in our halo catalogue. If the random number is less than or equal to $f_\mathrm{ON}(z)$, then that particular halo hosts a quasar. 

\begin{figure}  
  \begin{subfigure}[t]{0.49\textwidth}
    \centering
    \includegraphics[height=7.8cm, width=7.5cm]{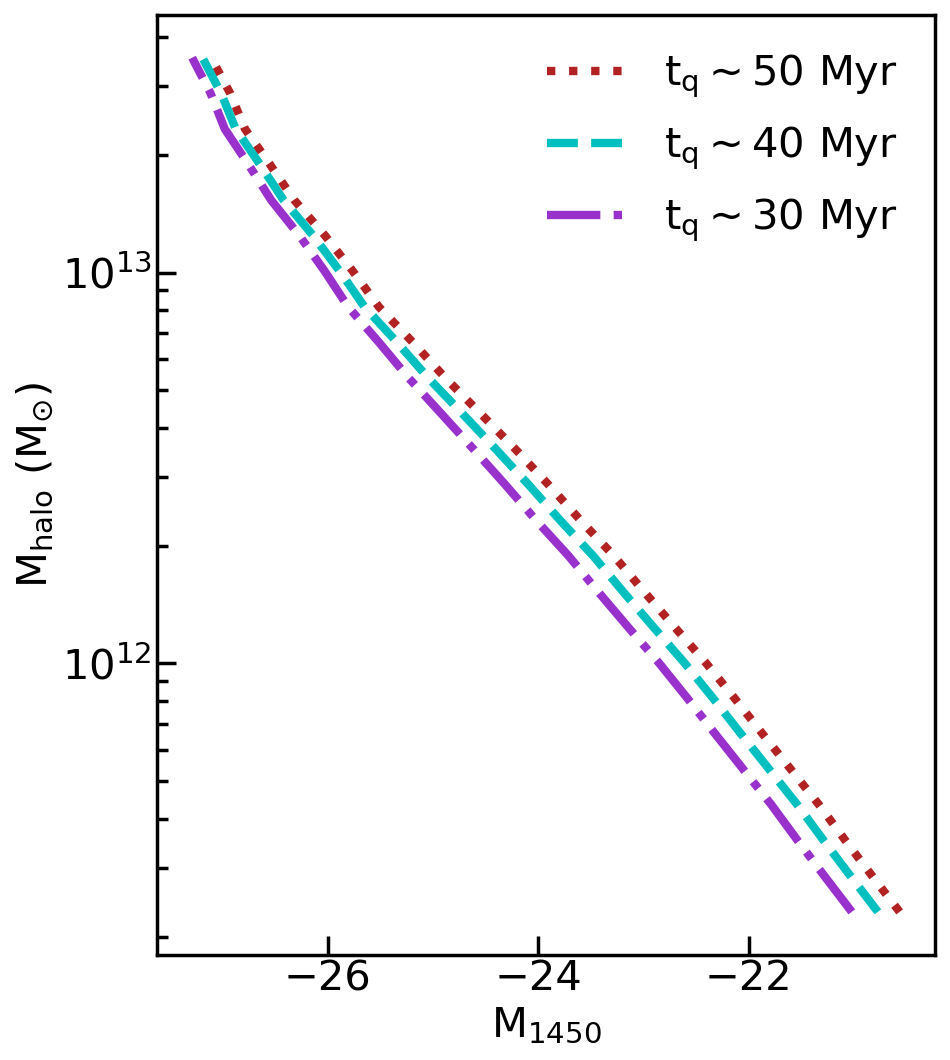}
    
  \end{subfigure}
  \hfill
  \begin{subfigure}[t]{0.49\textwidth}
    \centering
    \includegraphics[height=7.8cm, width=7.5cm]{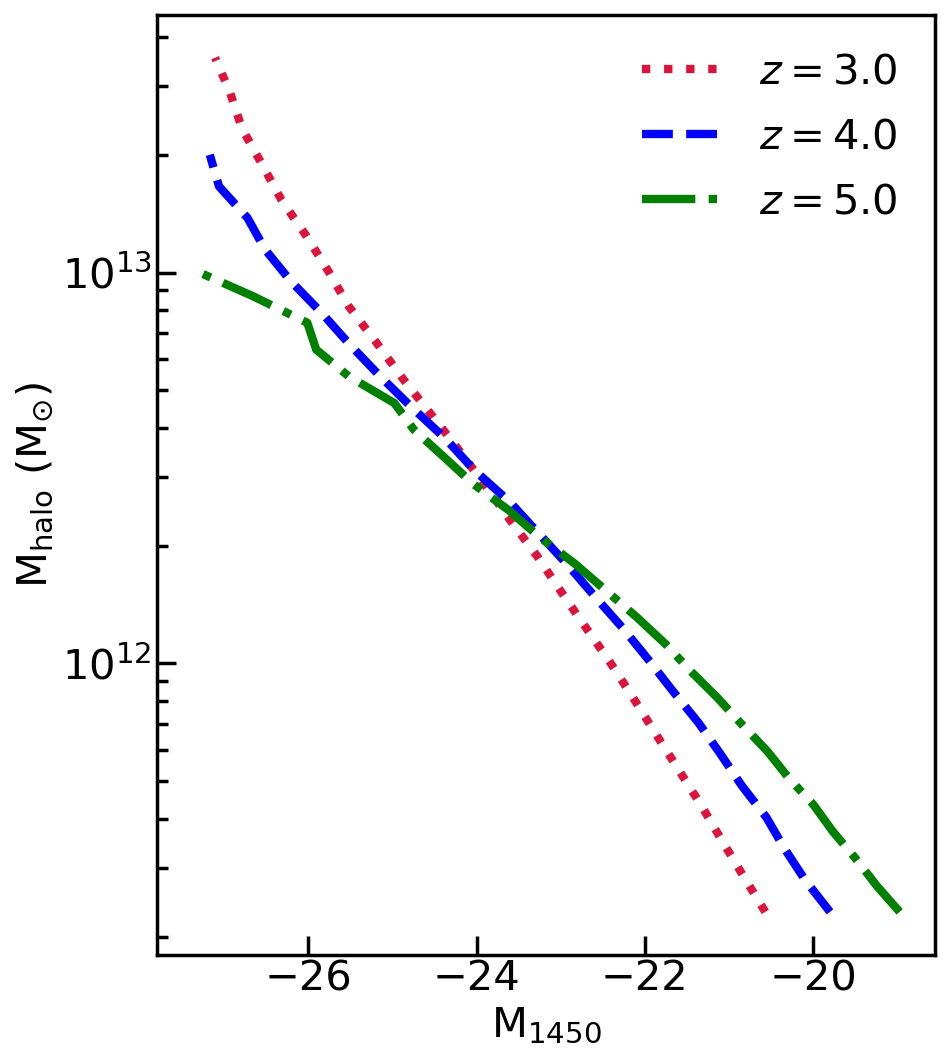}
   
  \end{subfigure}
  \caption{\textit{Left panel:} The relation between the halo mass and absolute magnitude (or luminosity) of quasars, for different quasar lifetimes at $z=3$. For a fixed value of the halo mass, a shorter lifetime maps to a brighter quasar. \textit{Right panel:} The relation between the halo mass and absolute magnitude for a quasar lifetime, $t_q = 40 \ \mathrm{Myr}$ at different redshifts.}
  \label{ML-ratio} 
\end{figure}

The result of our abundance matching procedure for different quasar lifetimes is given in the left panel of Figure~\ref{ML-ratio}. It is clear that for haloes within a fixed simulation box, a longer lifetime would map to fainter quasars than those with a shorter lifetime  for the same halo mass. In the right panel of Figure~\ref{ML-ratio}, we show the relation between the halo mass and quasar luminosity for our quasar model at different redshifts for a quasar lifetime of $40~\mr{Myr}$ . Given the monotonic relation between $\mr{M_{halo}}$ and $\mr{M}_{1450}$, the faintest magnitude for a given lifetime is limited by the resolution of the simulation box (or the least massive halo). Considering the shift towards brighter magnitudes with redshift, we fix a faint magnitude limit such that it is attainable at the lowest redshift of our interest, to avoid undercounting of fainter quasars at these redshifts due to finite box size. Therefore, we consider only quasars brighter than $\mr{M_{1450}} = -21$ to contribute to helium reionization. In subsequent sections, we shall find that in conjunction with the other free parameters of our reionization model, this choice leads to reasonable values of the average observed properties during helium reionization especially the ending at $z \sim 3$.

\subsection{Ionizing photons}
\label{sec:ion_ph}

Once the luminosity $\mr{L_{1450}}$ associated with a halo mass $\mr{M_{halo}}$ is obtained, we calculate the HeII ionizing photon emission rate for an assumed spectral energy distribution function. The rate of emmission of HeII ionizing photons from a quasar with luminosity $\mr{L_{1450}}$ is given by : 
\beq
\dot{N}= \int_{\nu_{He}}^{\infty} ~\cfrac{L_{\nu}}{h_P\nu} ~d \nu 
       = \cfrac{L_{\nu_{He}}}{h_P} \int_{\nu_{He}}^{\infty} \left( \cfrac{\nu}{\nu_{He}}\right)^{-\alpha_{UV}} \cfrac{1}{\nu}~d\nu  , 
       \label{qndot}
\eeq
where $\nu_{He}$ corresponds to the HeII ionizing threshold having a wavelength of $228$ \AA ~ and $\alpha_{UV}$ is the quasar spectral index which describes the spectral energy distribution(SED) of quasars, modelled to be a power-law ($f \propto \nu^{-\alpha_{UV}}$). The spectra of individual quasars can vary dramatically from one quasar to another hence their average spectrum is usually constrained from an average composite spectrum constructed from those of individual quasars. While at wavelengths greater than $912 \ $\AA~the SED is well constrained it shows a wide spread at smaller wavelengths.  The reported values of the SED index at these wavelengths are as low as $0.56$ \citep{scott2014} and 0.72 \citep{Tilton_2016} to as high as $1.57$ (or $1.96$) \citep{telfer2002rest} and $1.7\pm 0.61$ \citep{lusso}. Moreover, the smallest wavelength reached by these existing measurements is $\lambda \sim 400 \ $\AA. Therefore, the value of the SED index at wavelengths corresponding to the HeII ionizing wavelength, $\lambda \sim 25 \ $\AA~is extrapolated from these measured values which is reasonable if they reproduce the observables relevant for HeII reionization \citep[e.g.][]{Khaire_2017}. We use the following template from \cite{lusso} to define the SED:
\beq
L_{\nu} \propto \nu^{-\alpha}
\begin{cases}
& \alpha=0.61,\lambda > 912 \text{\AA} \\
& \alpha=\alpha_{UV}, \lambda \le 912 \text{\AA} ,
\end{cases}
\label{template}
\eeq
which leads to:
 \beq
L_{\nu_{He}}= L_{1450}  \left(\cfrac{228}{912}\right)^{\alpha_{UV}} \left(\cfrac{912}{1450}\right)^{0.61} ,
\eeq

Finally, from the above procedure we construct our quasar catalogues at redshift intervals separated by the quasar lifetime $t_q$ and retain only quasars having $\mr{M_{1450}} \le -21$. Our model for populating the dark matter haloes with quasars is described by two parameters, namely, (i) the quasar life time $t_q$ and (ii) the quasar spectral index $\alpha_{UV}$. In later sections we shall study the impact of these parameters on the progress of helium reionization. 

\section{Modelling Helium Reionization}\label{sec.3}

\subsection{Modified SCRIPT}
We model helium reionization in the \textbf{S}eminumerical \textbf{C}ode for \textbf{R}e\textbf{I}onization with \textbf{P}ho\textbf{T}on conservation (SCRIPT) which was originally developed for modelling hydrogen reionization \citep{Tirth_script}. The code  circumvents the problem of photon non-conservation \citep{Paranjape_2016} and associated  resolution dependence of the large scale 21-cm power spectrum in excursion set based semi-numerical models of reionization. This allows us to simulate reionization in a coarse grid thereby increasing the efficiency of our code. SCRIPT was recently extended to include inhomogeneous recombinations and thermal fluctuations \citep{Maity2022a} to obtain constraints on the parameter space during hydrogen reionization \citep{Maity2022b, Maity3}. 

In this work, we modify the base version of SCRIPT \footnote{\url{https://bitbucket.org/rctirthankar/script/src/master/ }}  to include inhomogenous helium reionization and obtain the corresponding thermal and ionization history during this epoch. We begin our simulation at a redshift of $z_{ini}=5.5$, assuming that all gas in the IGM is in the form of HII and HeII in ionization equilibrium. In order to generate ionization maps, an ionizing photon field and a matter density field is provided as an input to SCRIPT. Since the simulations are run at relatively coarse resolution, we can assume the baryonic densities to trace the dark matter. The total number of He atoms (which are initially all in the form of HeII) in a given grid cell $i$ is then given by:
  \beq
   N_{HeII,i}= \Delta_i \bar{n}_{He} V_{cell},
   \label{nhe}
   \eeq
where $\Delta_i = \rho_i/\bar{\rho}=1+\delta_i$ is the matter overdensity in a cell with volume $V_{cell}$ and $\bar{n}_{He}$ is the cosmic mean number density of helium.  We use the cloud-in-cell algorithm to generate the matter overdensity in the uniform coarse grid of SCRIPT from the dark matter particle positions in \texttt{GADGET4} snapshots described in Section \ref{sec.2}. The number of ionizing photons produced within a cell $i$ at a redshift $z_k$ within a time-step $\Delta t$ is given by:
\beq
   N_{ion,i} (z_k)=\sum_{j=1}^{\mathrm{No~of~quasars~in}~i} \dot{N}_{(j,z_{k-1})} \times \Delta t,
   \label{nph}
   \eeq
where $\dot{N}_{(j,z_{k-1})}$ is the rate of emission of ionizing photons for the $j$'th quasar (calculated using equation~\ref{qndot}) at the previous redshift snapshot $z_{k-1}$. The summation is carried out over all quasars within the cell $i$. Subsequently, the cumulative number of ionizing photons produced within a cell $i$ at a redshift $z_k$, denoted by $N^{tot}_{ion,i} (z_k)$ is computed by carrying out the summation over all photons produced in that cell starting from the initial redshift $z_{ini}$.

Given the ionizing photons and matter density field during helium reionization, we will now briefly outline the algorithm used in SCRIPT to generate ionization maps. In the absence of recombinations, a given cell in SCRIPT will absorb ${N_{HeII,i}}$ number of the total $N^{tot}_{ion,i} (z_k)$ photons produced by it. A central cell is flagged as completely ionized if the total number of ionizing photons produced in that cell are more than the number of HeII ions in that cell:
\beq
\cfrac{N^{tot}_{ion,i}(z_k)}{\bar{n}_{He} V_{cell}} \ge \Delta_i,
\label{frac}
\eeq
where we used eq. \ref{nhe}.
The excess remaining photons in the central cell are distributed to neighbouring cells in increasing order of distance until all its photons are exhausted. This means that the extra available photons are equally distributed to those cells which are located at the same distance from the central cell. For any given cell $j$ if the number of available ionizing photons is more than the number of HeII ions in that cell, then it is flagged as fully ionized, otherwise it is assigned an ionized fraction of $\mr{x}_{HeIII,j}={N^{avail}_{ion,j}}/{N_{He,j}}$. Since this process is carried out for each central cell independent of others, this can lead to 'overionized' cells having $\mr{x}_{HeIII,j} > 1$. Therefore, in a subsequent step the excess photons in these 'overionized' cells are again redistributed to surrounding partially ionized cells independent of each other. This step is carried out in multiple iterations until all such overionized cells are accounted for.  

The algorithm within SCRIPT ensures that sources emit radiation isotropically. However, in reality the ionizing UV emission from quasars would be anisotropic due to the dusty material surrounding the central supermassive black hole. Previously, the beaming of helium ionizing photons has been included in simulations by introducing angular emission \citep{Sokasian} or an obscuration factor \citep{Mcquinn2009}. Beaming may enhance features in the isolated ionized regions along the axis of the beam, especially at the beginning of reionization.  However, as the number of quasars increases, the average impact on the IGM at later redshifts is not expected to be much different from the isotropic case since an ionized region could be under the impact of several randomly oriented quasar beams. Thus, the average ionization and thermal properties of the IGM would not be affected majorly.

\subsection{Inhomogenous Recombinations}
\label{sec:inhomog_rec}
Due to fluctuations in the matter density field and rarity of quasars, helium reionization is expected to be a patchy process. In the following subsection we will also see that the temperature fluctuations resulting from helium reionization would depend upon inhomogeneous recombinations. Therefore, it is important to model inhomogeneous HeIII recombinations within SCRIPT which we implement analytically. 

Given the number density of HeIII ions $n_{HeIII,i}$ and electrons $n_{e,i}$ in a grid cell $i$, the recombination rate of HeIII to HeII is  given by:
\beq
\cfrac{d n_{rec,i}}{dt}=C_{HeIII,i} \times \alpha_{B,i} n_{HeIII,i} n_{e,i} \times(1+z)^3
\label{rec_rate}
\eeq
where $n_{rec,i}$ is the number of recombinations per unit comoving volume, $C_{HeIII,i}$ is the clumping factor which accounts for enhanced recombinations in regions with more number density of HeIII ions and $\alpha_B$ is the Case-B recombination coefficient. For helium reionization, the large ionized bubbles have sizes which are of the order of the mean free path of the hard photons, hence Case-B recombination coefficient is an appropriate choice \citep{Mcquinn2009}. 

The recombination coefficient has a mild dependence on the temperature of the grid cell and decreases with temperature. This dependence couples the ionization history of the grid cell with its thermal evolution described in the next subsection. This coupling allows for inferring the IGM thermal history during He reionization using the Ly$\alpha$ forest of hydrogen, as the temperature increase would also impact the HII recombination coefficient. The recombination rate per mean number of helium atoms is:
\beq
\cfrac{1}{\bar{n}_{He}}\cfrac{d n_{rec,i}}{dt}=C_{HeIII,i}\times \alpha_{B,i} \mathrm{x}_{HeIII,i} \bar{n}_e  \times(1+z)^3 \times \Delta_i^2,
\eeq
where $\bar{n}_e$ is the mean number density of electrons in the IGM \footnote{In an IGM with fully ionized hydrogen and doubly ionized helium, $\bar{n}_e= \Omega_b \rho_c \left(1-Y_P/2 \right) / m_p$, where $m_p$ is the proton mass and $\rho_c$ is the critical density of the universe.}. Since electrons from previous ionizations would also contribute to recombinations, the total number of recombinations per mean number of helium atoms at a given redshift:
\beq
\cfrac{n_{rec,i} (z)}{\bar{n}_{He}} =\int_{z_{ini}}^{z} dz \cfrac{dt}{dz}\times\cfrac{1}{\bar{n}_{He}}\cfrac{d n_{rec,i}}{dt}
\eeq
Due to inclusion of recombinations, a grid cell would require more photons to get ionized and therefore the condition in equation~\ref{frac} for a fully ionized central grid cell in SCRIPT becomes:
\beq
\cfrac{N^{tot}_{ion,i}(z_k)}{\bar{n}_{He} V_{cell}} \ge \Delta_i + \int_{z_{ini}}^{z} dz \cfrac{dt}{dz}\times\cfrac{1}{\bar{n}_{He}}\cfrac{d n_{rec,i}}{dt}
\eeq
We define the clumping factor $C_{HeIII,i}$ in a given grid cell using the globally averaged clumping factor $\mathcal{C}_{HeIII}$ \cite{Maity2022a}. The globally defined clumping factor is related to $C_{HeIII,i}$ for a mass weighted ionized fraction $Q^M_{HeIII,i}$ as:
\beq
\mathcal{C}_{HeIII}=\cfrac{\langle C_{HeIII,i} \ \Delta_i^2 \ \mathrm{x}_{HeIII,i} \ \rangle}{Q^M_{HeIII}} = \cfrac{\langle C_{HeIII,i} \ \Delta_i^2 \ \mathrm{x}_{HeIII,i} \ \rangle}{ \langle {\mathrm{x}}_{HeIII,i} \ \Delta_i \rangle} 
\label{eq:clumping}
\eeq

If we assume $C_{HeIII,i}$ to be the same for all grid cells, then one can infer $C_{HeIII,i}$ for each cell for a given $\mathcal{C}_{HeIII}$, using the above equation. In reality $C_{HeIII,i}$ would depend upon the subgrid density distribution \citep{Miralda_Escude_2000}. The global clumping factor $\mathcal{C}_{HeIII}$ is chosen to be a free parameter or a redshift dependent factor which we shall elaborate in later sections. 
This simplistic assumption of ignoring dependence of $C_{HeIII,i}$ on subgrid density distribution introduces a moderate resolution dependence on quantities which depend upon patchiness of reionization for the same value of $\mathcal{C}_{HeIII}$. However, as we shall see the dependence is weak compared to the uncertainties in our observable.

 The Case-B  HeIII recombination timescale at mean density at a temperature of $10^4 \ K$ at $z \sim 3$ is about 1 Gyr, substantially larger than the quasar lifetimes. This recombination timescale is even shorter at higher redshifts and densities. Since the cumulative effect of the ionizing radiation in a region, from previous redshifts is not as strong as that from a quasar, it will not be able to maintain the ionized fraction of HeIII against this recombination if the quasar turns off. Therefore, large ionized regions formed as a result of short-lived quasars can survive long after their source switches off \citep{Fossil}. We find that in our simulation grid the number of grid cells which are a part of such recombining regions increases gradually as reionization progresses and peaks when $\mr{x}_{HeIII} \sim 0.5$ where $\sim 30 \%$ of the grid cells are a part of such recombining regions after which their number drops abruptly.

\subsection{Photoionization Heating} \label{Sec:photo_heat}

The evolution of the kinetic temperature $T_i$ of a grid cell in an expanding universe is given by \citep[e.g.][]{Hui1997,Bolton2009}:
\beq
\cfrac{dT_i}{dz}=\cfrac{2T_i}{1+z}+\cfrac{2T_i}{3 \Delta_i} \cfrac{d \Delta _i}{dz}-\cfrac{T_i}{n_{tot,i}}\cfrac{dn_{tot,i}}{dz}+\cfrac{2 \epsilon_i}{3 k_B n_{tot,i}} \cfrac{dt}{dz},
\label{temp}
\eeq
where $n_{tot,i}$ is the total number of baryons in the grid cell, $k_B$ is the Boltzmann constant and $\epsilon_i$ is the heating (or cooling) rate per unit comoving volume.
The first term in the above equation arises due to the adiabatic expansion of the universe while the second term because of the adiabatic compression and expansion due to structure formation. The third term accounts for the change in temperature due to change in the number of particles. The last term appears due to other heating and cooling processes each described by a separate $\epsilon$. Other processes include processes which can heat the IGM, such as photoheating from the ionizing photons or processes which cool the IGM like recombination cooling, collisional cooling, free-free cooling and cooling due to Compton scattering of electrons off CMB photons. We find that the impact of all the cooling processes are negligible at the temperatures, densities and the redshift regime that we are working in (for the values of the cooling coefficients taken from \citep{Hui1997}). We assume that all photons above the ionizing threshold of HeII contribute to ionization and hence amount to photoheating. In actual, the more energetic photons would not contribute substantially to ionization in the immediate vicinity of the sources. Therefore, these photons heat up the IGM almost uniformly further away from the sources \citep{Sanderbeck2020}.
The first two terms in equation~\ref{temp} can be directly computed from the overdensity values of each grid cell for the given redshift snapshot. The third term due to change in total number of particles can be written as:

\beq
\cfrac{T_i}{n_{tot,i}}\cfrac{dn_{tot,i}}{dz} = T_i \left[2 \left(\frac{4 - 3 Y_P}{Y_P} \right) + x_{\mathrm{HeIII},i}\right]^{-1} \cfrac{d\mr{x}_{HeIII,i}}{dz},
\eeq
where we have used $n_{tot,i}=n_{H,i}+n_{He,i}+n_{e,i}$ as all of the hydrogen is singly ionized and helium is either in the form of HeII or HeIII.

The temperature evolution due to photoheating during helium reionization is described by the following expression:
\begin{align}
\cfrac{d T_i}{dt} &= 2 \left( \cfrac{2}{Y_P} - 1 \right) T^{He}_{re} C_{HeIII,i} \alpha_B^{HeIII} \mathrm{x}_{HeIII,i} \ n_{He,i} (1+z)^3 + T^{He}_{re}\cfrac{d \mathrm{x_{HeIII}}}{dt}
\nonumber \\
& \quad +T^{H}_{re} C_{HII,i} \alpha_A^{HII} n_{H,i} (1+z)^3 , 
 \label{temp_ev}
\end{align}
 where $T_{re}^{H}$ and $T_{re}^{He}$ is the reionization temperature for HI and HeII respectively. It describes the temperature boost corresponding to the excess energy per photoionization (of HI or HeII), distributed over all baryons. The value of $T_{re}^{He}$ is taken to be a free parameter while $T_{re}^{H}$ is fixed to a value of $1.85 \times 10^4 \ K$ obtained from constraints during hydrogen reionization \citep{Maity2022b}. Note that we use the Case-A recombination for hydrogen since hydrogen is assumed to be fully ionized and hence optically thin to ionizing photons. The global clumping factor for hydrogen, $\mathcal{C}_{HII}$ is derived from the fits from hydrodynamical + N-body simulations \citep{Shull}:
\beq
\mathcal{C}_{HII}=2.9 \left( \cfrac{1+z}{6} \right) ^{-1.1}
\label{C_H}
\eeq 
There are other fitting formulas for the mean clumping factor \citep[e.g.][]{Pawlik,Finlator}, which give values of the same order at our redshifts of interest.
The full derivation of equation~\ref{temp_ev} is given in Appendix~\ref{B}. The equation has been derived assuming that our large grid cells can be thought of as consisting of neutral and ionized regions. Thereafter, the ionization of neutral regions is assumed to take place in a timescale that is short compared to the Hubble timescale and HeIII recombination timescale. The ionized parts of the grid cell are assumed to be in photoionization equilibrium between simulation time steps. These assumptions hold because the recombination timescale and the Hubble timescale are both large compared to the time step of $\sim 40$ \rm{Myr} of our simulations. The first term in equation~\ref{temp_ev} arises from the fully ionized HeII in the grid cell in ionization equilibrium. The second term is the temperature increase due to HeII regions getting ionized. The last term arises because of fully ionized hydrogen in ionization equilibrium.  In case of the relic ionized regions which are recombining actively, the temperature evolution is only dictated by the first three terms in equation~\ref{temp}.

Since we start our simulations in a regime where hydrogen is fully ionized, the initial temperature of each pixel would have a temperature corresponding to the heating due to preceding hydrogen reionization. The temperature and density asymptotes to a tight power-law relation in the low-density photoionized IGM after hydrogen reionization ends  \citep{Hui1997}. The tight power-law arises due to the form of the temperature dependence of the recombination coefficient ($\alpha_A(T) \propto T^{-\beta}$) which leads to the photoheating rate per $n_H^{2/3}$ of a gas parcel to be nearly independent of $n_H$ \citep{Mcquinn2016}. Realistically, since different regions ionize, heat and cool at different timings we do not expect to obtain a single power-law relation. However, measurements of the temperature density equation of state from quasar spectra rely on fitting the temperature density scatter to a single power law \citep{Prakash2020}.  Therefore, each cell in our simulation is assigned an initial temperature of $T_i=T_0 \Delta_i ^{  \gamma-1}$ at $z_{ini}$, where $T_0$ is the temperature at mean density and $\gamma$ is the slope of the power law. We use empirical measurements of $T_0$ and $  \gamma$ \citep{Prakash2020} at $z_{ini}=5.5$ to set initial temperature of each grid cell. Since the scatter between temperature and density decreases during the end stages of reionization \citep{Maity2022a}, it is a good approximation to assume a single power law relation. 

\begin{figure}  
  \centering
  \includegraphics[height=14.5cm,width=15.48cm]{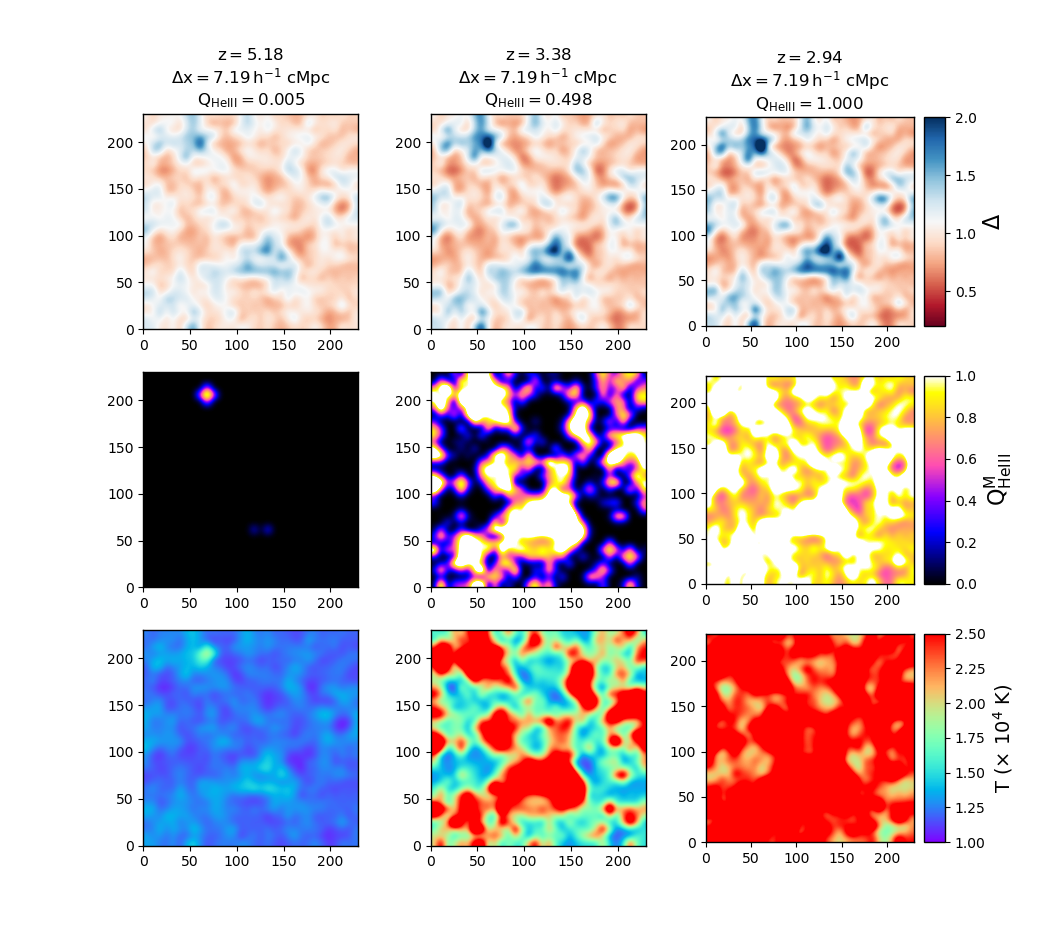}
  
\caption{\textit{Top to bottom:}. Maps for overdensity $\Delta$, mass averaged ionized fraction $Q^M_{HeIII}$ and temperature $T$ generated in our $230$ $\mathrm{Mpc ~ h^{-1}}$ box, at three different redshifts: $z \sim 5.2$ (\textit{Left}), $z \sim 3.25 $ ($\textit{Middle}$) and $z \sim 2.80$ ($\textit{Right}$). These maps are generated on a $32^3$ grid resolution in SCRIPT.}
   \label{maps}
\end{figure}

\section{Results}\label{sec.4}
In order to demonstrate the performance of our simulation to probe helium reionization, we shall fix a \textit{fiducial} model and infer the ionization and thermal history such that it fits two main observables: firstly, the ending of helium reionization at $z \sim 3$ and secondly, the most recent measurements of the temperature density equation of state  by \cite{Prakash21} during these redshifts.

We assume that quasars have a universal lifetime of $t_q \sim 40 \ \mathrm{Myrs}$ and the simulation time-step is $\Delta t = t_q$ or shorter.  Only quasars brighter than $\mr{M_{1450}} \le -21$ are assumed to contribute to helium reionization. With these assumptions our \textit{fiducial} model is described by three main free parameters: (1) The quasar SED index - $\alpha_{UV} = 1.7$ (2) The global clumping factor for HeIII - $\mathcal{C}_{HeIII} = 15.6$ (3) The temperature increase per unit baryon per HeII ionization - $T^{He}_{re} = 6000 \ K$. In Figure~\ref{maps}, we show the maps for overdensity ($\Delta$), mass averaged ionized fraction ($Q^M_{HeIII}=\mr{x}_{HeIII} \Delta$) and temperature ($T$) for our \textit{fiducial} model in a $32^3$ grid at three different stages of helium reionization. The choice of grid resolution is motivated by the fact that it is the lowest resolution beyond which our simulations converge relative to the errors in the observations (Appendix~\ref{Converge}). From the figure, we see that unlike the case for hydrogen reionization, not every high density region hosts a source, this is especially visible at $z=5.18$, where helium reionization is just beginning. Secondly, while the ionized and heated regions correlate with high density regions, they are slightly larger and more connected than the underlying density field. This is due to the large sizes of ionized bubbles around quasars, which are able to travel far beyond their sources residing in high density regions and therefore also ionize and heat up the voids in the immediate vicinity. 

\begin{figure}  
  \begin{subfigure}[t]{0.49\textwidth}
    \centering
  \includegraphics[height=7.8cm,width=7.5cm]{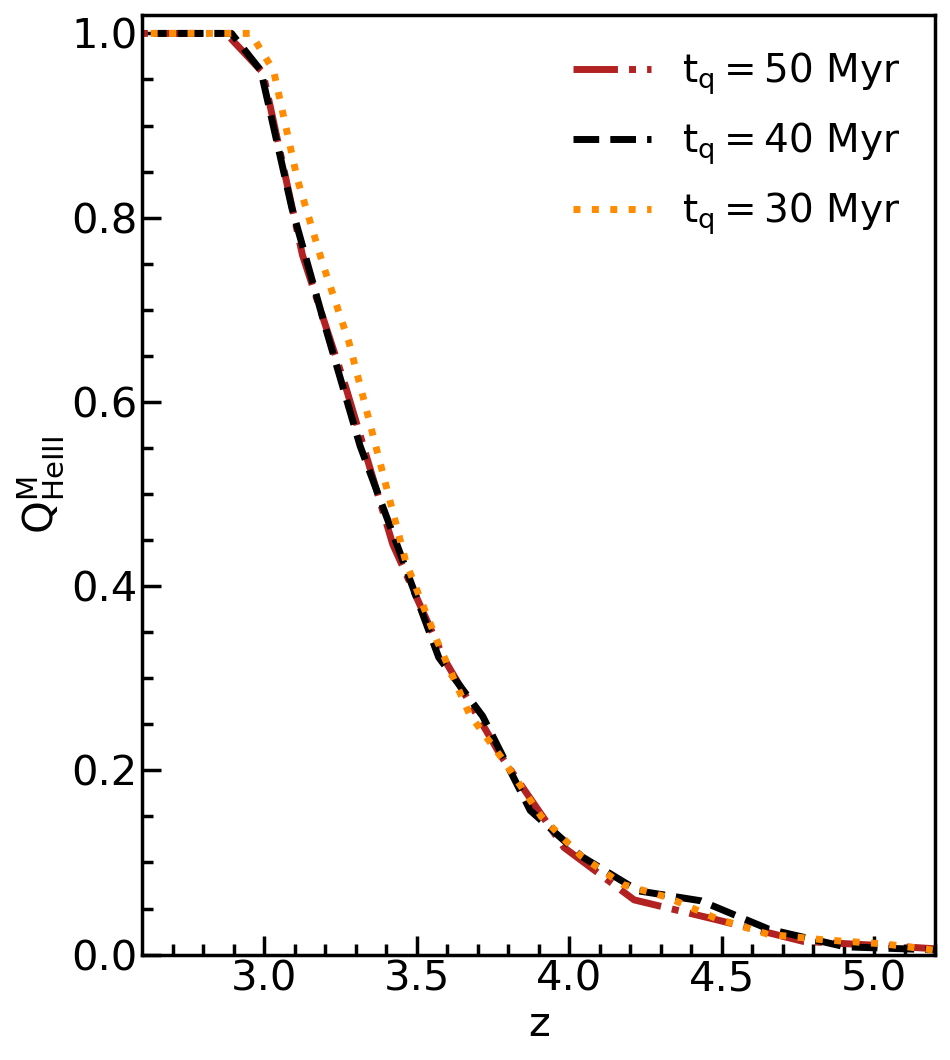}
  \end{subfigure}
  \hfill
  \begin{subfigure}[t]{0.49\textwidth}
  \centering
  \includegraphics[height=7.8cm,width=7.5cm]{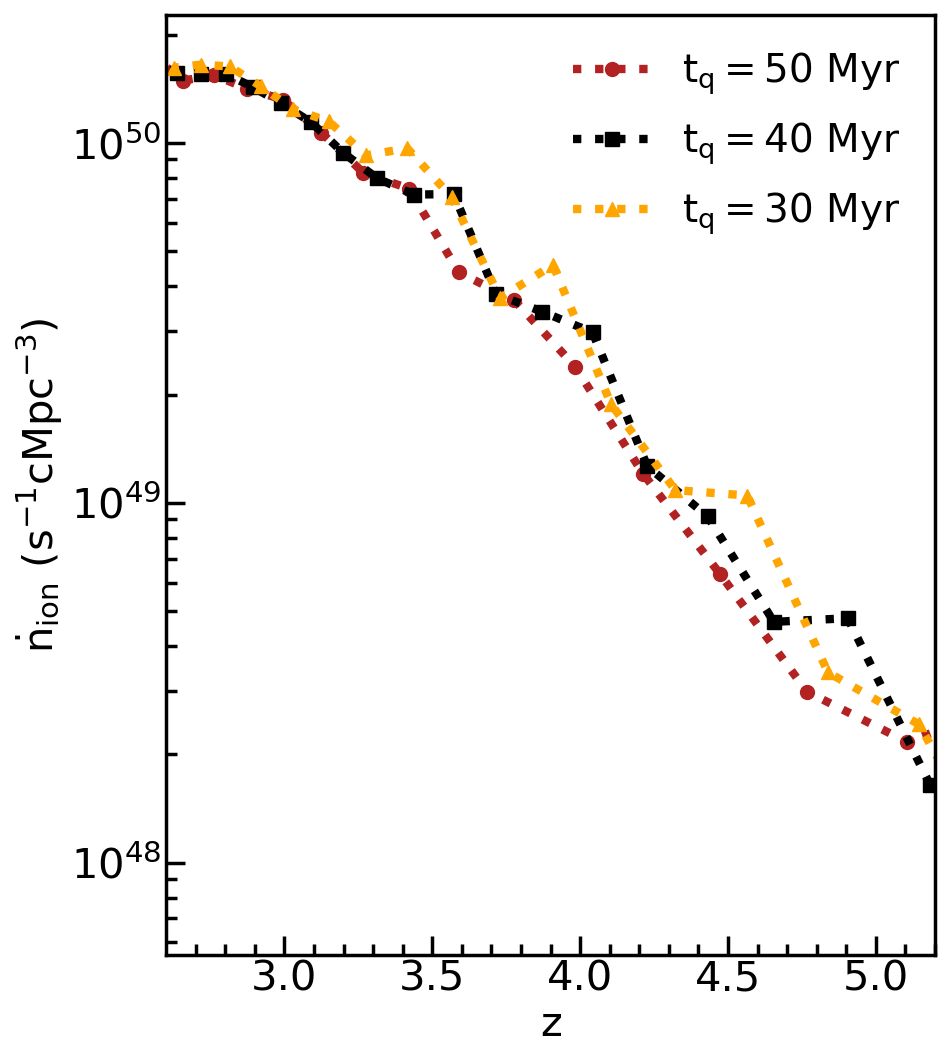}
  \end{subfigure}
  \caption{Mass averaged ionization history for our \textit{Fiducial} model for different quasar lifetimes (\textit{Left}) and variation of corresponding rate of ionizing photon emission with redshift. Only quasars brighter than $\mathrm{M_{1450} \le -21}$ are assumed to contribute to helium reionization in this model.}
  \label{ion_hist_tq}
  \end{figure}

\subsection{Ionization History}

\begin{figure}  
  \centering
  \includegraphics[height=7.8cm,width=7.5cm]{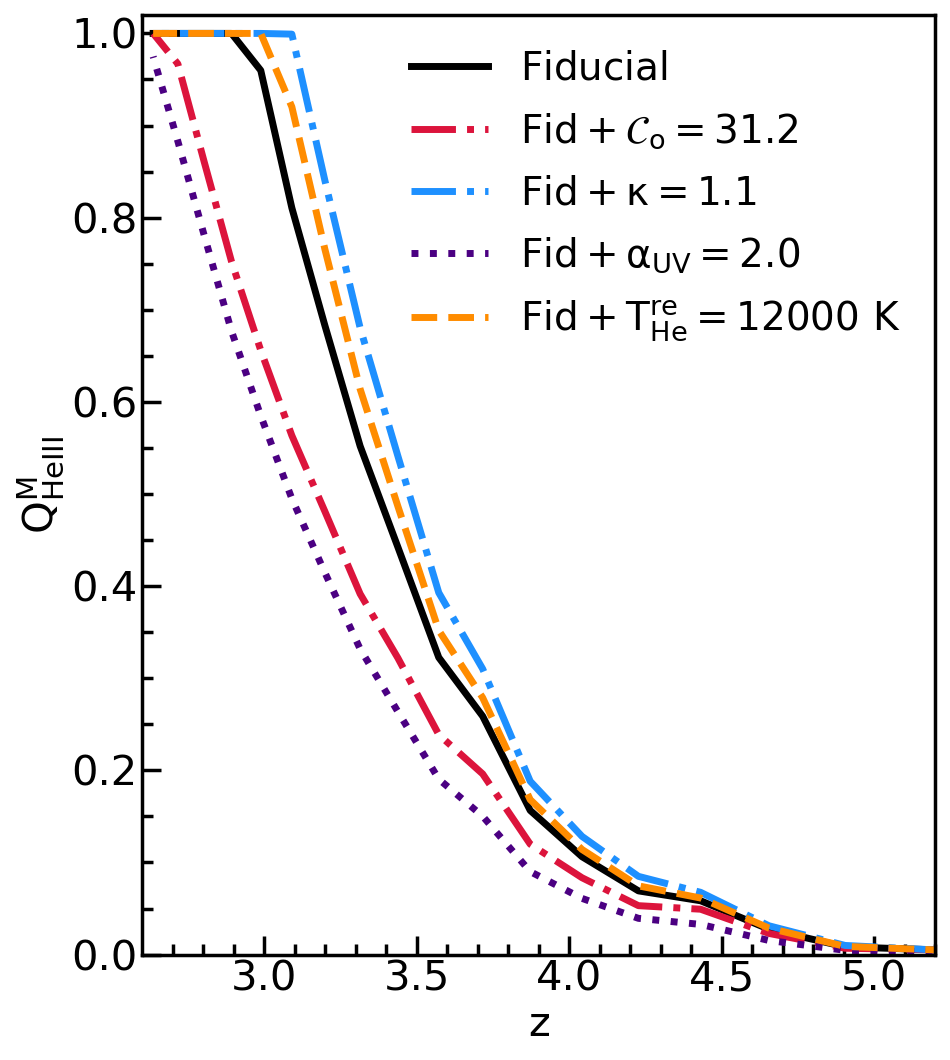}

  \caption{Ionization history for different combinations of our free parameters. The fiducial model is plotted in black solid line and is the model with $\mathcal{C}_{HeIII}=15.6 \ (\mathcal{C}_0=15.6, \ \kappa=0), \ \alpha_{UV}=1.7 \ and \ T^{re}_{He} = 6000 \ K$. The other models show the effect of varying the values of each parameter if all the other parameters are fixed to those for the fiducial model. The red and the blue dot dashed lines show the effect of changing $\mathcal{C}_0$ and $\kappa$ to $31.2$ and $1.1$ respectively while other parameters are fixed to the fiducial values. The purple dotted line plots the model with $\alpha_{UV}=2$ and the yellow dashed line with $T^{re}_{He}=12000 \ K$ with clumping factor fixed to the fiducial value.}
  \label{ion_hist_mod}
\end{figure}

Before we proceed to study the impact of varying our free parameters about the \textit{fiducial} model, it would be interesting to see the impact of varying the quasar lifetime $t_q$ on the ionization history during helium reionization. The left panel of Figure~\ref{ion_hist_tq} shows the mass averaged ionized fraction $Q^M_{HeIII}$ for three different quasar lifetime values feasible within our box for quasars brighter than $\mr{M_{1450}}=-21$ (see right panel of Figure~\ref{ML-ratio}). While quasar lifetimes are expected to alter the topology of ionized regions, from the figure we find that there is no significant effect on the ionization history. This is because we are assuming a universal quasar lifetime and abundance matching for the same number density of haloes would lead to fewer long-lived quasars or a larger number of short-lived quasars, resulting in similar total number of ionizing photons. This can be seen in the right panel of Figure~\ref{ion_hist_tq} which shows the evolution of the rate of mean photon emission per comoving volume ($\dot{n}_{ion}$) at a given redshift for different quasar lifetimes. The fluctuations in $\dot{n}_{ion}$ are a result of having discrete sources in our simulation box.

We will now elaborate upon the effect of varying the free parameters of our model. First, we show the effect of varying the clumping factor $\mathcal{C}_{HeIII}$. Since clumping factor describes the enhanced recombination rate in denser regions, a higher clumping factor would slow down reionization. In principle the clumping factor must vary with redshift, since the IGM becomes clumpier at lower redshifts. We define a power law variation for clumping factor, analogous to \cite{Shull} as described in equation~\ref{C_H}:
\beq
\mathcal{C}_{HeIII}=\mathcal{C}_0 \left( \cfrac{1+z}{3} \right) ^{- \kappa},
\label{C_He}
\eeq
This would introduce two more free parameters in our model such that for our fiducial model $\mathcal{C}_0=15.6$ and $\kappa=0$. However, as shown in Figure~\ref{ion_hist_mod} (red and blue dashed dotted line) the effect of an increase in $\kappa$ is same as that of decrease in $\mathcal{C}_0$. The values of the other three parameters are set to those for the fiducial model which is plotted as the black solid line. An increase in $\mathcal{C}_0$ by two times the clumping factor of the fiducial model ($\mathcal{C}_0=15.6$) delays the ending of reionization to $z \sim 2.6$. On the other hand, setting the redshift evolution similar to that used for hydrogen, i.e.  $\kappa=1.1$ for $\mathcal{C}_0=15.6$ speeds up reionization. 

The second free parameter in our model is the quasar spectral index $\alpha_{UV}$. An increase (decrease) in $\alpha_{UV}$ implies a softer (harder) quasar spectrum, leading to emission of less (more) number of ionizing photons. Therefore, the impact of increasing (decreasing) $\alpha_{UV}$ is to shift the end of reionization to a later (earlier) redshift as shown by the purple dotted line in Figure~\ref{ion_hist_mod}. The value of $\alpha_{UV}=2$ leads to relatively softer spectra compared to that for the fiducial model having a value of $\alpha_{UV}=1.7$ plotted in black.

The last free parameter of our model is $T_{re}^{He}$. This would impact the temperature of the IGM and hence the recombination rate coefficient  which shows a decrease with increasing temperature ($\alpha_B(T) \propto T^{-0.62}$). An increase in temperature leads to a lower recombination rate resulting in an earlier end to reionization. A larger value of $T_{re}^{He}$ relative to the fiducial model would lead to a higher temperature for the same number of HeII ions ionized. The impact of varying $T^{re}_{He}$ to a value which is twice as high as the  fiducial value of $T^{re}_{He}=6000 \ K$ on the ionization history of helium is plotted as the yellow dashed line in Figure~\ref{ion_hist_mod}. Since the temperature dependence of the recombination coefficient is mild, the impact of varying $T^{re}_{He}$ on the ionization history is not very large.

\subsection{Thermal History}

 We first describe the general picture of the thermal evolution during helium reionization, as expected from equation~\ref{temp_ev} substituted as the fourth term in equation~\ref{temp}. Initially, before helium reionization commences only the third heating term in equation~\ref{temp_ev} and the two adiabatic terms in equation~\ref{temp} would determine the thermal evolution. Expansion of the universe would lead to cooling, while structure formation in high (low) density regions would lead to heating (cooling) of the IGM. The photoionization heating due to helium reionization, would alter the asymptotic equation of state which was established after the preceding epoch of hydrogen reionization. The IGM would again settle down to another power-law relation after helium reionization ends \citep{Sanderbeck2016}. 

In reality, different regions would get ionized at different times and therefore follow the above evolution at different rates. This introduces a scatter around the mean temperature density equation of state which reduces after a reionization event ends. The above implementation of the thermal history in large grid cells like ours, does not capture the temperature evolution of the low density IGM which is relevant for inferring the thermal history from Ly$\alpha$ forest measurements of quasar spectra. Additionally, we did not include baryons in our dark matter only simulation due to which any interpretation of gas temperature from our simulation is incomplete.  In order to incorporate the two effects and retain the speed of our simulations, we implement a sub-grid approach to model the low density IGM in a more realistic fashion. The approach is similar to that used in \citep{Maity2022a}, but since we are in a lower redshift regime, we account for mildly non-linear sub-grid densities. Additionally, since during helium reionization, some main grid cells are actively recombining, we also account for sub-grid recombinations and allow for partially ionized sub-grid fractions. Our approach is decribed below:

\subsubsection{Subgrid Implementation of thermal history}

\begin{enumerate}
    \item The main effect of baryons on a given dark matter density field is to smooth it at the Jean's scale. In order to incorporate this effect, at the starting redshift $z_{ini}$ we generate a lognormal distribution of $N_{sub}$ sub-grid density elements within each main grid cell. The elements are distributed such that their mean is equal to the value of the density contrast of the main grid element, $\delta_i$ \footnote{Since our grid cells have a resolution that is much larger than the Jean's scale at the redshifts of interest, the baryonic overdensity traces the dark matter overdensity at these scales. Therefore, it is reasonable to assume that the mean of the sub-grid baryonic density elements is equal to the density contrast of the main grid cell, even though it is generated from dark matter only simulations.}. The standard deviation $\sigma(R)$ used for generating the distribution is calculated at a scale $R$ equal to the comoving Jeans scale at that redshift \citep{Binney}: 
    \beq
    R_{Jeans}=\left( \cfrac{c_s^2}{4 \pi G \Omega_m \rho_c (1+z)} \right)^{1/2},
    \eeq
    where $G$ is the Gravitational constant, $c_s=\sqrt{\cfrac{\gamma_p k_B T}{m_p}}$ is the sound speed and we take the polytropic index $\gamma_p=5/3$ for a monoatomic gas.
    We calculate $\sigma(R)$ using PyCCL package \footnote{\url{https://github.com/LSSTDESC/CCL}} with the HALOFIT non-linear power spectrum \citep{halofit}. 
    
    \item Equipped with the initial subgrid density contrast elements, we next evolve the densities at each redshift. In order to do that we first map each density element at $z_{ini}$ to a corresponding linear density contrast using the following parametric formulae from \citep{M01996}:
    \beq
1+\delta_{NL} =\cfrac{9}{2} \cfrac{(\theta - \mathrm{sin}~\theta)^2}{(1-\mathrm{cos}~\theta)^3} \ \ , \ \ \ \delta_L=\cfrac{3 \times 6^{2/3}}{20} (\theta -\mathrm{sin}~\theta)^{2/3}
\eeq
for $\delta_{NL} > 0$,

\beq
1+\delta_{NL} =\cfrac{9}{2} \cfrac{(\mathrm{sinh}~\theta - \theta)^2}{(\mathrm{cosh}~\theta -1)^3} \ \ , \ \ \ \delta_L=-\cfrac{3 \times 6^{2/3}}{20} (\mathrm{sinh}~\theta - \theta)^{2/3}
\eeq
for $\delta_{NL} < 0$.

These equations are numerically inverted to obtain a mapping between $\delta_L$ and $\delta_{NL}$. 

     Since we are only concerned with the thermal history of the low density IGM, it is safe to use the fitting formula. These mapped densities are then evolved linearly, such that $\delta^{sub}_L (z) = D(z) \delta^{sub}_L(z_{ini})$, where $D(z)=\cfrac{1+z_{ini}}{1+z}$ is the linear growth factor. These linearly evolved densities are then mapped back to the non-linear densities using the same fitting formula. 
 
    \item Next, we track the ionization history of these subgrid elements. For a given main grid cell at a redshift $z$ if there is an increase in its ionized fraction by $d\mathrm{x_{HeIII,i}}$ at the next redshift step, we assign randomly selected subgrid elements an ionization fraction of $1$ until the total increase of $d\mathrm{x_{HeIII,i}}$ is achieved. Owing to a high recombination rate of HeIII, we also find main grid elements for which there is a decrease in the ionization fraction over the timestep of our simulation. These grid cells are a part of relic ionized regions \citep{Fossil} which were described at the end of Section ~\ref{sec:inhomog_rec}. Therefore, for such regions we incorporate recombinations, such that the rate of change of ionization fraction due to recombinations:
    \beq
    \cfrac{d \mathrm{x}_{sub}}{dt} = -\alpha_B(T_{sub}) \  \mathrm{x}_{sub} \ \bar{n}_e \ \Delta_{sub} \ (1+z)^3 
    \eeq
    This follows from equation~\ref{rec_rate}, since an increase in number of recombinations by  $dn_{rec}$ will lead to a corresponding decrease in number of HeIII ions by $d n_{HeIII}=-dn_{rec}$. Here, we assume a clumping factor of $C_{HeIII,i}=1$ as there is no further substructure within such regions.
     Thus, the ionization fraction of each subgrid element after a timestep $\Delta t$ from the above equation: 
     \beq
    \mathrm{x}_{sub}(t+\Delta t) \simeq \mathrm{x}_{sub}(t) ~ \mathrm{exp}[-\alpha_B \bar{n}_e \Delta_{sub}(1+z)^3 \Delta t],
    \label{relic_rec}
    \eeq
    where it is assumed that the densities do not evolve considerably over a time-step $\Delta t$ and also that the change in $\alpha_B(T)$ due to change in temperature is negligible \footnote{Since, $\cfrac{d \alpha_B(T)}{dt} \propto \cfrac{1}{T^{1.62}} \cfrac{dT}{dt}$ and the temperatures are of the order of $10^4$ K.}. 
Since we only have a knowledge of the ionization fraction of our subgrid elements and we do not know how many photons each subgrid element would absorb, we cannot directly incorporate recombinations by integrating equation~\ref{rec_rate} over all previous redshifts.
 This approximation for decreasing the ionization fraction to incorporate recombinations is similar to that used in \cite{Gleser2005}.  Finally, the ionization fraction of all subgrid elements is adjusted mathematically such that the mean of their ionized fractions matches the ionization fraction of the main grid. If the ionization fraction of the main grid is $\mr{x_{HeIII}}$ and the mean of  $N_{sub}$ subgrid elements within it is $\mr{\bar{x}}_{sub}$, then the required change in ionization fraction of $N_1$ of those subgrid elements for both the ionization fractions to be equal is: 
 \beq
 d\mr{x}_{sub}=\cfrac{N_{sub}}{N_1} (\mr{x}_{HeIII,i}-\bar{\mr{x}}_{sub})
 \eeq
 Since the approximation in equation~\ref{relic_rec} does not account for recombinations from previous redshifts, this re-adjusment of the mean implicitly incorporates the unaccounted recombinations. Also, this mathematical re-adjustment fixes any numerical offset in regions where $d \mathrm{x_{HeIII,i}}\ge 0$. Due to this, some subgrid elements may attain an ionized fraction $\mathrm{x_{sub}} < 1$ instead of $1$ in such regions. Thus, by construction the mean of the ionization fraction of all subgrid elements $\bar{\mr{x}}_{sub}$ \textit{exactly} matches the ionization fraction of a given main grid cell $\mr{x}_{HeIII,i}$.
    
    \item Thermal history of each subgrid element is tracked by solving the temperature evolution equation~\ref{temp_ev} by assuming a clumping factor of 1 for both hydrogen and helium since these subgrid elements are homogeneous without any further substructure. In grid cells corresponding to the relic regions there is no heating since the only dominant process in these regions is recombinations as was described in Section~\ref{Sec:photo_heat}.
 \end{enumerate}
\begin{figure}  
  \centering
  \includegraphics[height=10.cm,width=9.cm]{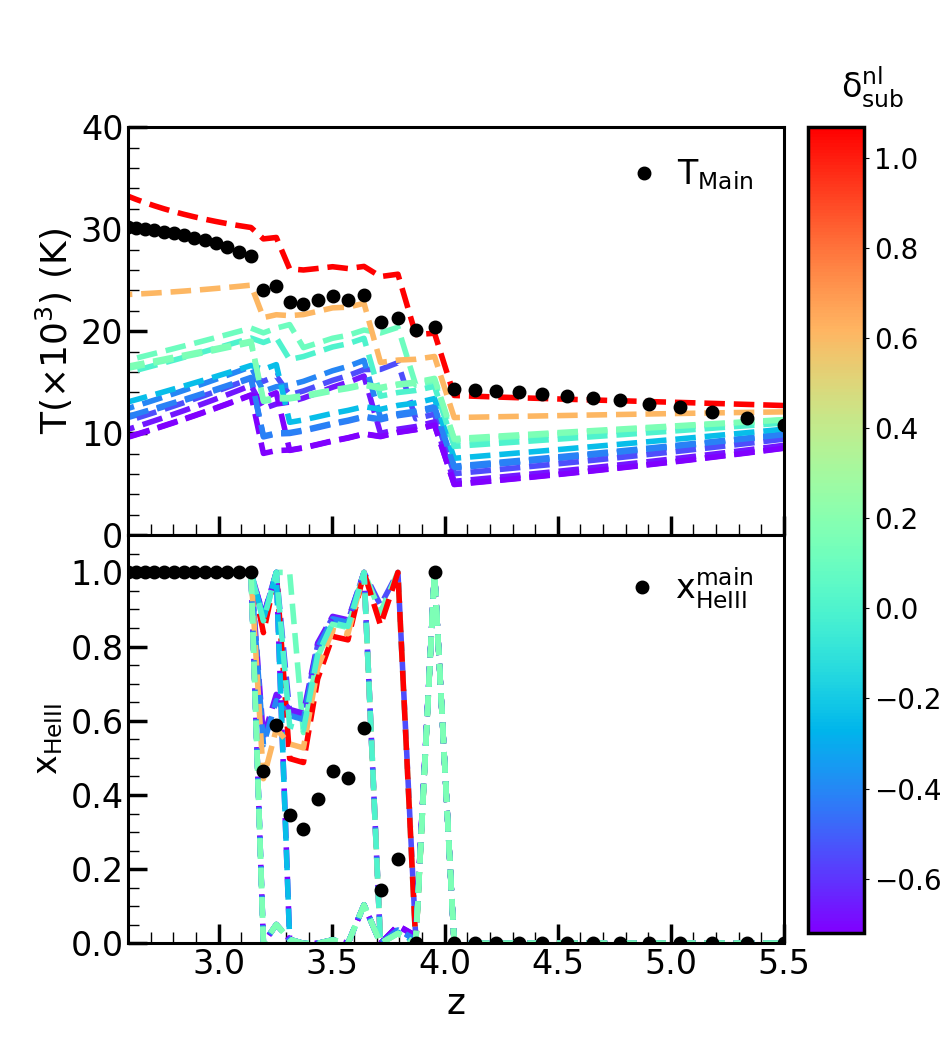}
  
  \caption{The 15 subgrid elements inside a single main grid element. The colorbar denotes the initial non-linear density contrast of the subgrid elements and the black dots shows the corresponding value of $T_0$ and $\gamma$ of the main grid cell.}
  \label{subgrid}
\end{figure}

\begin{figure}  
  \centering
  \includegraphics[height=6.cm,width=15. cm]{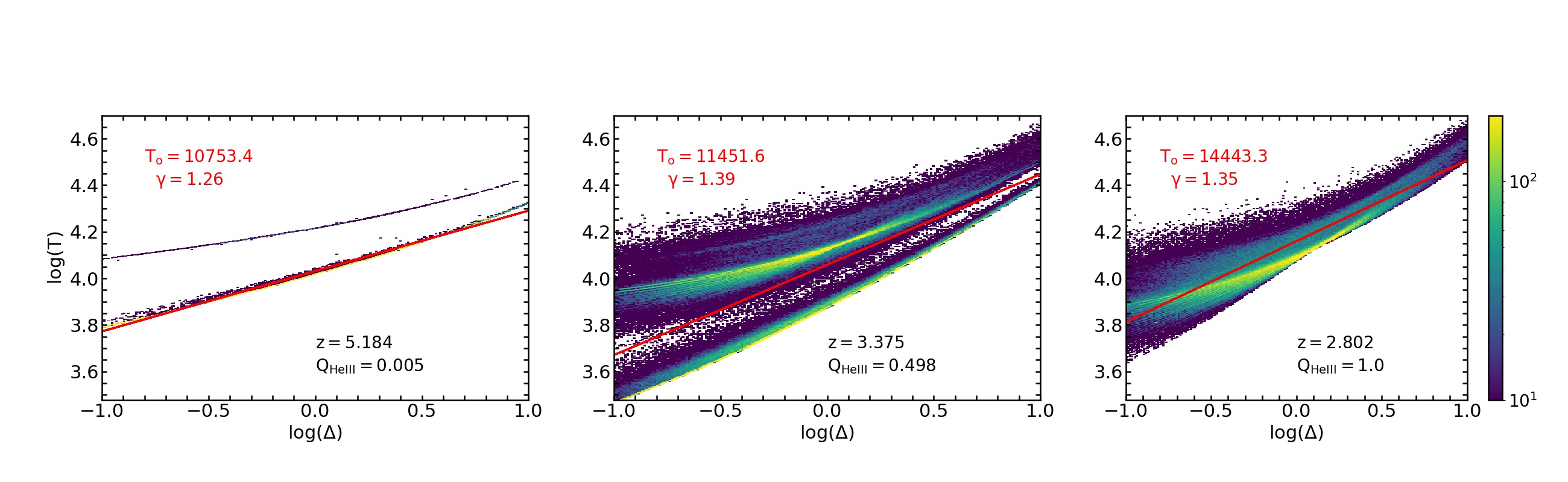}
 \includegraphics[height=6.cm,width=15. cm]{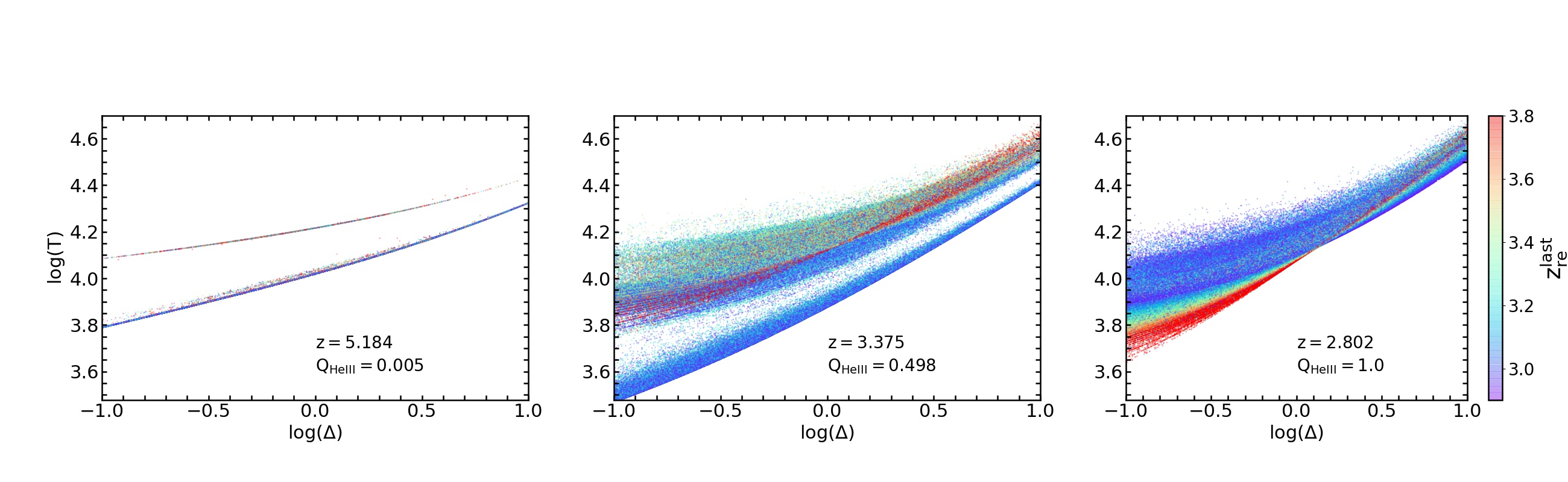}
  \caption{The top row shows the temperature density histogram for our subgrid elements at three different stages of reionization. The colorbar denotes the number density of points in a given two dimensional bin of the histogram. The lower row shows the corresponding scatter plot, where the colorbar denotes, $z^{last}_{re}$ the lowest redshift at which a subgrid element gets finally ionized.}
  \label{sub_scatter}
\end{figure}

In order to incorporate subgrid thermal history within our box, we use $N_{sub}=15$ elements in our thermal history model. Our model converges with respect to the number of subgrid elements (see Appendix~\ref{Converge}). In Figure~\ref{subgrid} we show the evolution of temperature (top panel) and ionization fraction (bottom panel) of the subgrid elements within a main grid cell. The black dots are the values for the main grid cell and the dashed lines show the corresponding 15 subgrid density elements color coded according to their density contrast value at $z_{ini}$. In the lower panel, we show how subgrid elements are able to physically capture recombinations for different values of the subgrid densities i.e. high density regions experience more recombinations than lower density regions during the redshift range when the region recombines (at $3.3 \le z \le 3.7$). Additionally, the figure also shows how randomly selected elements get ionized when there is a positive change in the ionization fraction of the main grid (for example the first peak at $z \sim 4$ and the red and the blue peaks at the subsequent time steps, after which they start recombining). The corresponding subgrid temperature evolution is plotted in the upper panel. At low redshifts, once helium reionization is complete in all subgrid elements($z \lesssim 3$), only the first and third term in equation~\ref{temp_ev} along with the adiabatic evolution terms in equation~\ref{temp} determine the thermal history. For larger density values the increase in temperature due to adiabatic collapse and heating due to ionization equilibrium leads to an overall increase in temperature. On the other hand for low density regions, it is the adiabatic cooling term that dominates, leading to an  overall cooling. A similar evolution in cooling while all subgrid elements are still neutral is also seen at $z>4$ except that heating is only due to the ionization equilibrium of HII and HeII (third term in equation~\ref{temp_ev}), since helium has not yet ionized. Since the recombination rate is higher in high density regions, there would be more ionizations required to maintain the photoionization equilibrium and hence more photoionization heating. This leads to higher temperatures in higher density regions when they are in photoionization equilibrium. Since only HII and HeII are in photoionization equilibrium, at these high redshifts the temperature evolution therefore follows a density trend which is similar to but less pronounced than that at low redshifts ($z \lesssim 3$).  Also note the black dots which describe temperature evolution of the main grid element. When compared with the evolution of a density element having similar density (the colored dashed line closest to the black point at $z=5.5$), the above expected evolution is not observed. This shows that the main grid cells are too coarse to study the thermal history evolution. 

In the top panel of Figure~\ref{sub_scatter} we show a 2D histogram of temperature versus overdensity for three different stages of HeII reionization color coded by the density of points.  As helium reionization progresses, the plots show that at a given density there exists a bi-modality in temperature. This bi-modality is typical of helium reionization seen in hydrodynamic simulations \citep{Compostella_2013, Sanderbeck_2018} and semi-analytical techniques \citep{Gleser2005,Furlanetto_2008} and is a result of the patchiness of the reionization process. The lower temperature branch is the set of points which are yet to get ionized. At $z=z_{ini}$, all points are expected to lie on the straight line that follows the power law after hydrogen reionization, with which the initial temperature of the subgrid elements is set. In our case we set it similar to the main grid elements i.e. each subgrid element has an initial temperature of $T=T_0 \Delta_{sub}^{  \gamma -1}$, where $T_0=11000 \ K$ and $  \gamma=1.2$ at $z_{ini}=5.5$, set by the measurements of Lyman alpha forest transmission spikes \citep{Prakash2020}. Initially, all points would lie on the lower branch, the slope of which increases gradually with redshift, owing to the increase (decrease) in temperature due to adiabatic compression (expansion) of high (low) density regions and additional heating due to photoionization equilibrium of HII and HeII. There would also be a density independent decrease in temperature of this branch, due to adiabatic expansion of the universe. Gradually, as helium reionization begins some of the elements in the lower branch would heat up due to photoionization heating, leading to the temperature bi-modality. As the top left panel shows, initially we find that most of the points lie on the lower branch, since most of the helium is in the form of HeII. As helium reionization progresses, some of the subgrid elements experience photoheating which eventually increases the density of points in the upper branch of the scatter plot (top middle panel). Finally, as helium reionization ends, the bi-modality no longer exists as all subgrid elements have experienced photoheating due to helium reionization (top right panel) . We describe this evolution by fitting a straight line (red solid line) described by a power law $T=T_0 \Delta^{\gamma-1}$, where $T_0$ is then the temperature at mean density. The fit is obtained by binning $\mathrm{log~\Delta}$ between -1 to 1 at an interval of 0.125 and then fitting the mean temperature in each bin by a straight line. In the top panel of Figure~\ref{sub_scatter} we notice how the fitted line shifts to the upper temperature branch and its scatter increases as helium reionization is underway. As helium reionization completes, most points lie around the fitted single power law. The panels in the second row of Figure~\ref{sub_scatter} show the scatter plots corresponding to the same redshifts as the upper row, but instead color coded by the redshift $z^{last}_{re}$ at which a subgrid element \textit{finally} ionizes. Since a sub-grid element can belong to a relic ionized region it may ionize completely after more than one episodes of recombination and ionization (see Figure~\ref{subgrid}). Therefore, $z^{last}_{re}$ is the lowest redshift at which it ionizes and does not recombine again. The bottom right panel shows that while regions which get ionized earlier are cooler at lower densities, the trend is not so well-defined at higher densities. This is because, once completely ionized, the photoheating due to ionization equilibrium in HeIII regions would occur at a rate $t_{rec}$ which becomes smaller than the rate of cooling due to adiabatic expansion $t_{ad}$ at these temperatures and densities. Since adiabatic cooling is the only dominant cooling mechanism, this well-defined trend is not observed at these densities and temperatures. For example, $t_{rec} \lesssim t_{ad}$ at $\Delta \gtrsim 1$ for elements having $T=10^4 \ K$ and at $\Delta \gtrsim 2.4$ for  elements having $T=10^{4.5}$. Additionally, since most of these densities would also experience adiabatic heating due to structure formation (since their values are $\Delta \gtrsim 1$), the overall trend at these densities is that of increase in temperature, once they get fully ionized \footnote{For comparison in case of HII, $t_{rec} \lesssim t_{ad}$ at $\Delta \gtrsim 5.7$ for $T=10^4 \ K$}. This trend is also visible in Figure~\ref{subgrid} at $z \lesssim 3.0$ (e.g. the red dashed line), where we see an overall heating in high density regions. Therefore, a given high density element that gets ionized earlier can be hotter than an element with the same density that ionizes later, especially since other cooling mechanisms are all sub-dominant at these temperatures and densities. 
   \begin{figure*}
     \begin{subfigure}[t]{0.48\textwidth}
  \centering
     \includegraphics[width=\linewidth]{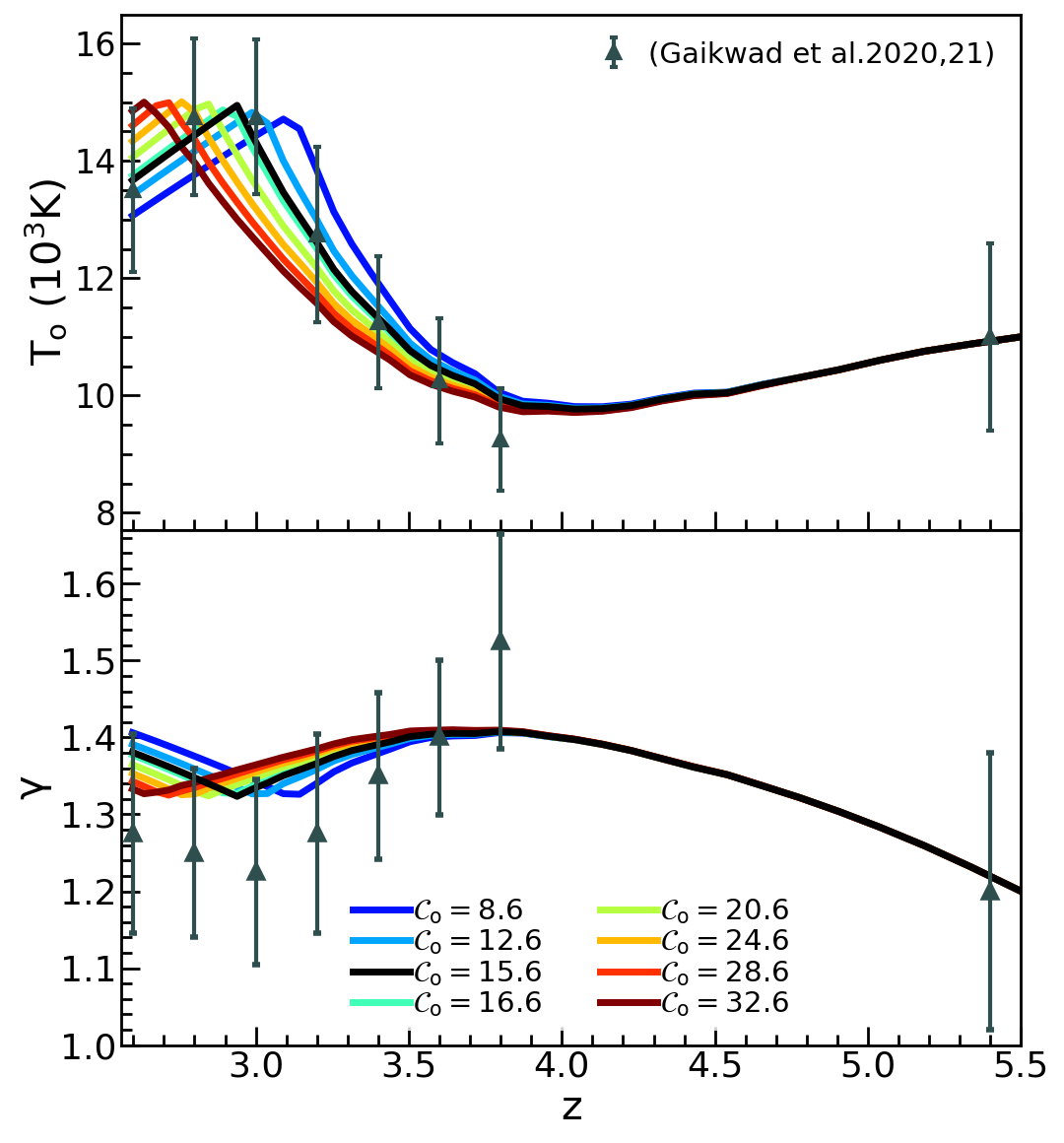}

\end{subfigure}
\hfill
  \begin{subfigure}[t]{0.48\textwidth}
  \centering
 \includegraphics[width=\linewidth]{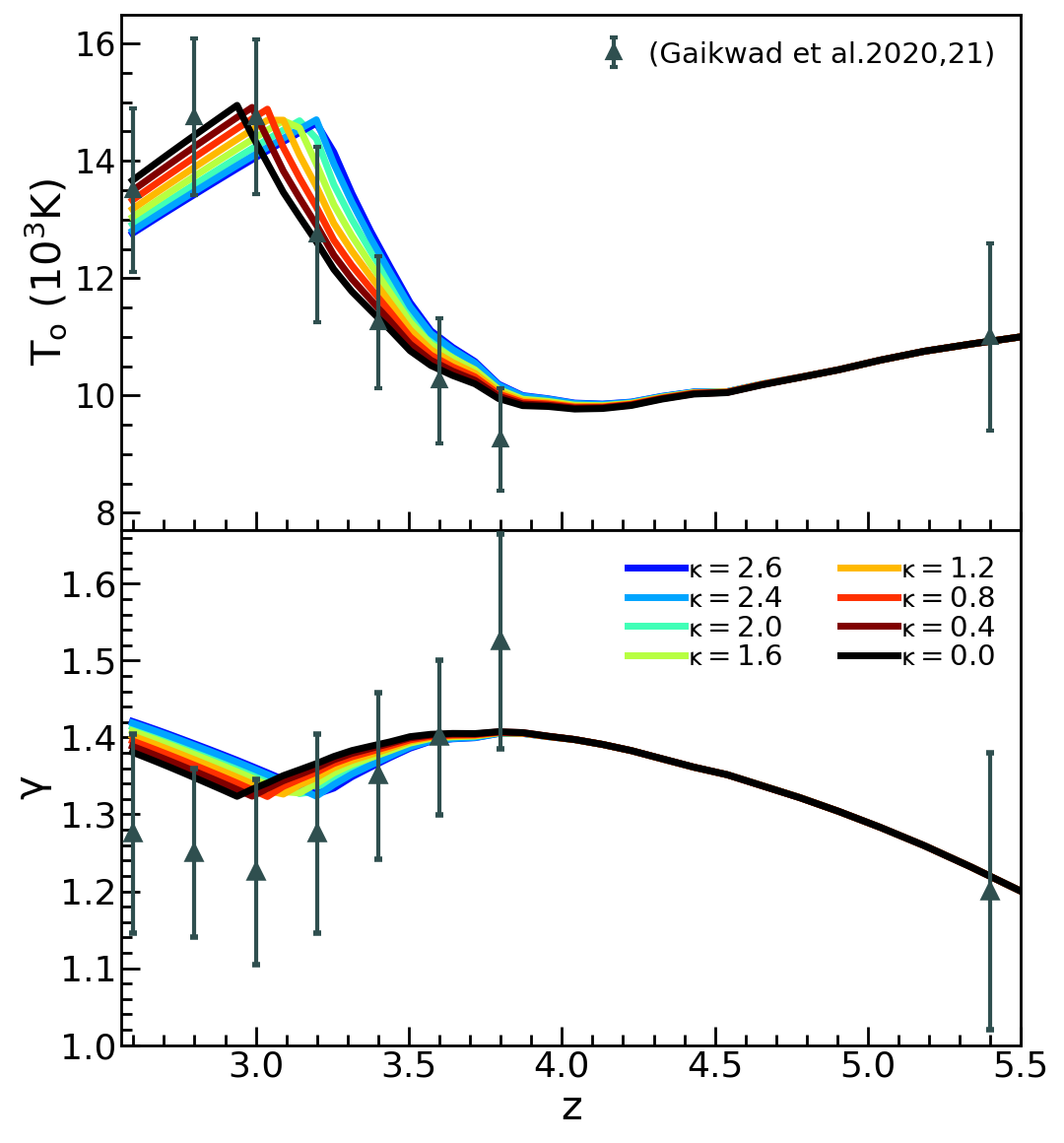}
 \end{subfigure}
\caption{The figure shows the variation of the amplitude $T_0$ and   slope $  \gamma$ of the temperature density equation of state for different values of the clumping factor as parameterized by equation~\ref{C_He} and $\alpha_{UV}=1.7$ and $T^{re}_{He}=6000 \ K$. The black solid line corresponds to the \textit{fiducial} value of $\mathcal{C}_0=15.6$ and $\kappa=0$.  The effect of clumping factor is to shift the temperature peak due to corresponding shift in the ending of reionization, with higher clumping shifting it to lower redshifts. The grey errorbars are the measured values from \citep{Prakash2020,Prakash21}.}

\label{fig:clump}
  \end{figure*}

  \begin{figure*}
     \begin{subfigure}[t]{0.48\textwidth}
  \centering
     \includegraphics[width=\linewidth]{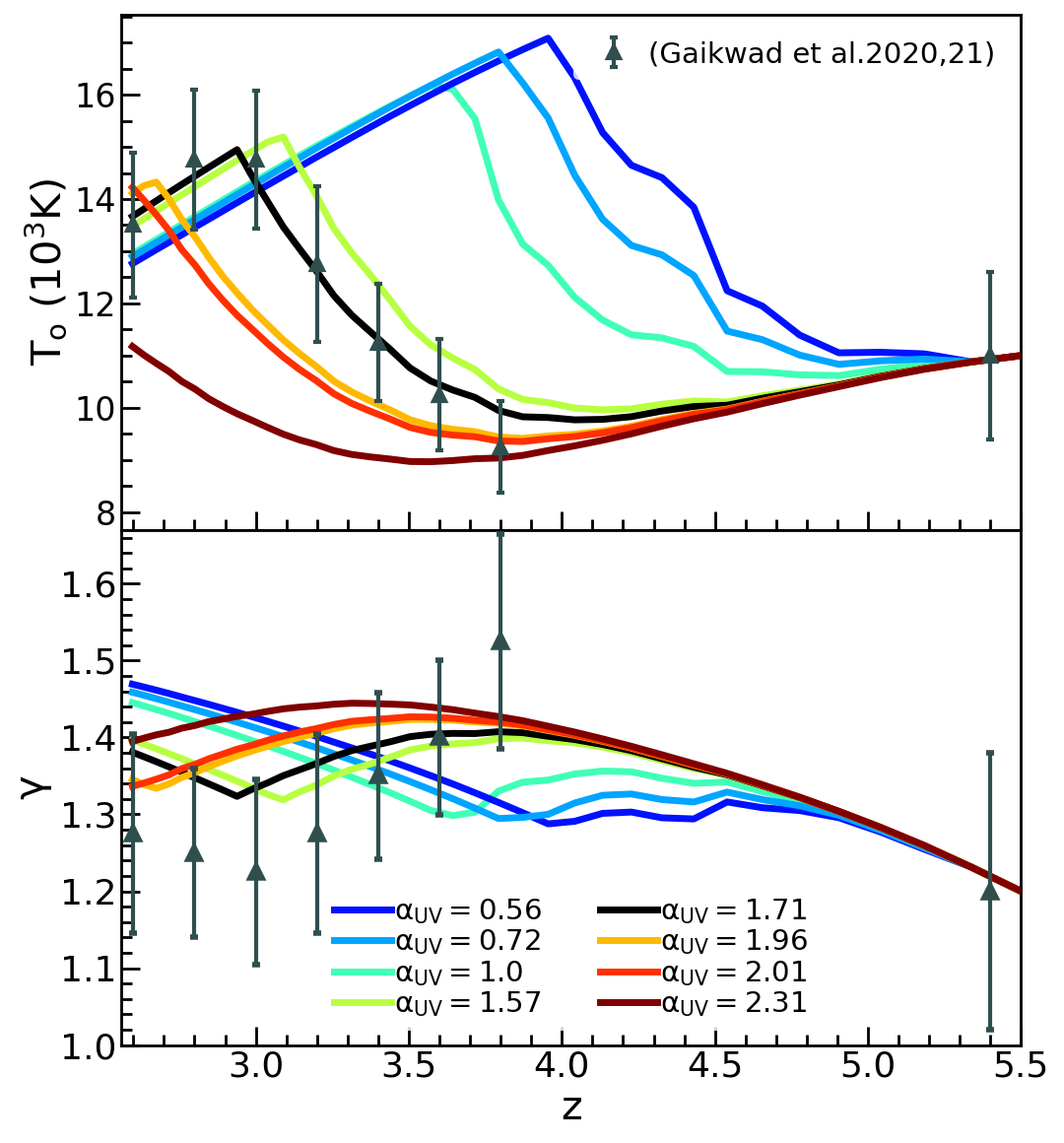}

\end{subfigure}
\hfill
  \begin{subfigure}[t]{0.48\textwidth}
  \centering
 \includegraphics[width=\linewidth]{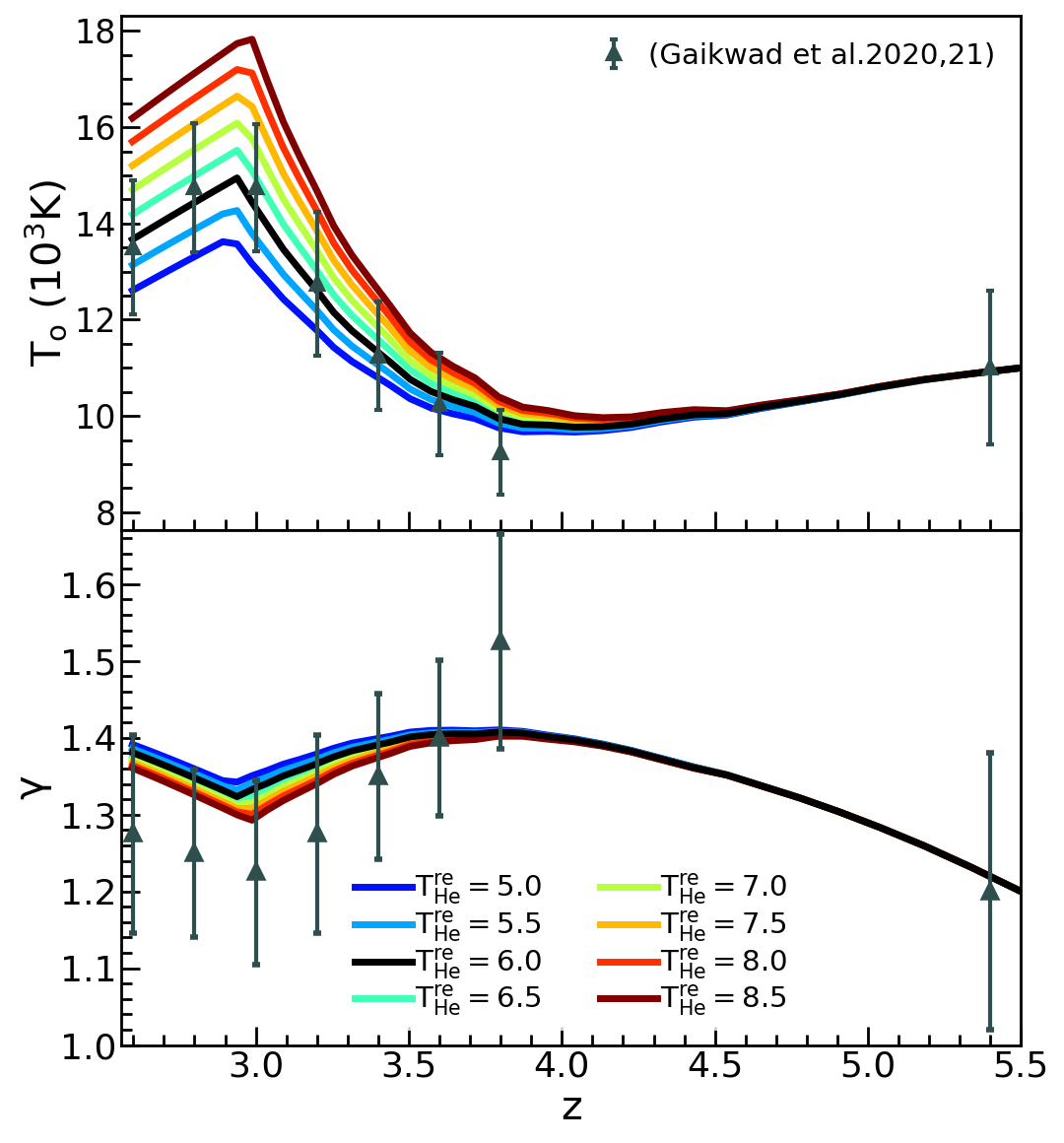}
 \end{subfigure}
\caption{The figure shows the variation of the amplitude $T_0$ and $  \gamma$ of the temperature density equation of state for different values of the quasar SED index \textit{(Left)} and the average temperature increase per HeII ionization, $T^{re}_{He}$ \textit{(Right)} for a fixed clumping factor of $\mathcal{C}_0=15.6$ and $\kappa=0$. The black lines in both the plots are those corresponding to the \textit{fiducial} model. In the left panel the different values of $\alpha_{UV}$ are the estimates from literature (see text) and in the right panel the values of $T^{re}_{He}$ have units of ($10^{3}~K$). The grey errorbars are the measured values from \citep{Prakash2020,Prakash21}.} 

\label{fig:alpha_tr}
  \end{figure*}
Next, we study the impact of varying our free parameters on the mean temperature-density equation of state obtained as above. Before helium reionization commences, the photoionized gas after hydrogen reionization begins to predominantly cool adiabatically until it is photoheated again during helium reionization. After helium reionization, this photoheated gas again predominantly cools. This leads to an initial decrease (increase) in $T_0$ ($  \gamma$) followed by an increase (decrease) when helium reionization picks up. The ending of helium reionization is marked by a sharp peak in the $T_0$ evolution and a corresponding dip in $\gamma$ evolution. In Figure~\ref{fig:clump} we show the evolution of $T_0$ and $\gamma$ for different values of $\mathcal{C}_0$ and $\kappa$. Since clumping factor only impacts the redshift of ending of reionization and has no impact on the amount of photoheating, the impact of increase (decrease) in $\mathcal{C}_0$ ($\kappa$) is to shift the position of the peak in $T_0$ (or dip in $\gamma$) to a later redshift.  In Figure~\ref{fig:alpha_tr} we show the variation with $\alpha_{UV}$ in the left panel and $T^{re}_{He}$ in the right panel. Harder quasar spectra leads to faster reionization and a much hotter IGM because the enhanced ionization leads to more photoheating. This increases the amplitude of $T_0$ and shifts its peak to earlier redshifts for a lower value of $\alpha_{UV}$. The different values of $\alpha_{UV}$ used for the plot are those which are measured in literature (see section \ref{sec:ion_ph}). On the other hand an increase in $T^{re}_{He}$ leads to a greater rise in temperature therefore increasing the peak value of $T_0$. Mild temperature dependence of the recombination coefficient $\alpha_B(T)$ also introduces a mild shift in the location of the peak as $T^{re}_{He}$ varies.
The grey errorbars in both the plots are the measured values of $T_0$ and $\gamma$ at $z=5.5$ from transmission spikes of quasars \citep{Prakash2020} and from Ly$\alpha$ forest statistics \citep{Prakash21} (between $z=2.6$ and $z=3.8$). The black solid line is the \textit{fiducial} model in both the plots. 

So far we fitted only the mean of the temperature in different density bins, however one can fit a different binned statistic like the median or fit the scatter plot directly. Ideally the choice of binned statistic would depend upon the corresponding statistics used for inferring the temperature density equation of state from observations with which we would compare our results. However, in our case since there is a well defined temperature bi-modality, fitting the mean instead of the median (Appendix~\ref{App:fit}), serves as a better measure of the equation of state. We have tested the convergence of our code with respect to resolution, time and number of subgrid elements relative to existing measurements of the temperature density equation of state. We also show the impact of choosing a different redshift of starting our simulations using these measurements and find that the redshift of ending of helium reionization is independent of the choice of $z_{ini}$. These convergence tests are presented in Appendix~\ref{Converge}.  

\section{Summary and Discussion}\label{sec.5}

The epoch of helium reionization is the last major baryonic phase transitions of the universe that impacts the ionization and thermal state of the IGM between $2.5 \lesssim z \lesssim 5$. This has important implications for using the Ly$\alpha$ forest for precision cosmology as the impact of helium reionization on the IGM can bias these measurements. Particularly, the HeII Ly$\alpha$ forest is an important probe of the low density IGM owing to its higher optical depth \citep{croft1997intergalactic}.
To this end, we have presented fast semi-numerical simulations of helium reionization by modifying the \textbf{S}eminumerical \textbf{C}ode for \textbf{R}e\textbf{I}onization with \textbf{P}ho\textbf{T}on conservation (SCRIPT) \citep{Tirth_script}. We extended SCRIPT by incorporating inhomogenous recombinations and an analytical model of thermal evolution of the IGM during helium reionization in a $\mathrm{230 \ h^{-1} \ Mpc}$ box. We used abundance matching to model quasars as sources of helium reionization by assigning luminosities to dark matter haloes generated from N-body simulations. The photon conserving nature of SCRIPT allows us to model reionization in a coarse grid leading to a substantial gain in efficiency. However, that renders our thermal history model incomplete since our grid resolution is large (for example $\sim \mathrm{7.2 \ Mpc}$ in our \textit{fiducial} model). Therefore, we model thermal history using a sub-grid prescription without compromising with the efficiency of our code. Our semi-numerical model then leads to four main free parameters -- the clumping factor (described by $\mathcal{C}_0$ and $\kappa$), the quasar SED index ($\alpha_{UV}$) and the average temperature increase per HeII ionization ($T^{re}_{He}$). Our \textit{fiducial} model with $\mathcal{C}_0=15.6$, $\kappa=0$, $\alpha_{UV}=1.7$ and $T^{re}_{He}=6000 \ K$ is able to reproduce the latest measurements of the temperature density equation of state described by $T_0$ and $\gamma$.  The combined effect of the variation of total clumping factor $\mathcal{C}_{HeIII}$ and $T^{re}_{He}$, is degenerate with the effect of varying $\alpha_{UV}$. Our sub-grid thermal model converges well with respect to time and resolution. When run on a $32^3$ grid, covering the evolution of HeII over 38 redshift snapshots between $z=5.5$ to $z=2.6$ our code runs in $\sim 30$ seconds for our \textit{fiducial} model.

However, it is to be noted that the $T_0-  \gamma$ measurements are obtained by calibrating against hydrodynamic simulations with some ionizing UVB model as input and is not a direct observable for helium reionization \citep[e.g.][]{Prakash2020,Prakash21}. The direct observable for inferring the end stages of helium reionization is the effective optical depth and the statistical measurements of HeII Ly$\alpha$ forest along different quasar sightlines \citep{Worseck19,Makan_2021,Makan_2022}. In that direction, there are still some limitations in our model which would especially become relevant if we use the measurements of the Ly$\alpha$ forest to constrain parameters with our semi-numerical framework. The Ly$\alpha$ forest lines are sensitive to small scale unionized HeII regions which would not be captured by our low resolution grid. Additionally, while our sub-grid model is able to reproduce the average thermal evolution of $T_0$ and $  \gamma$, the baryonic prescription is still approximate. A more accurate baryonic prescription, for example one which is calibrated against hydrodynamical simulations would be useful if the efficiency of our model is to be retained while making it more physically accurate. Lastly, we have assumed that only the UV photons which ionize the gas, contribute to heating i.e. we have ignored the thermal impact of large mean free path photons from quasars which can impact the heating in voids away from sources \citep[e.g.][]{Sanderbeck2020}. Also our simulation does not capture the pressure smoothing which is degenerate with the thermal broadening of Ly$\alpha$ forest lines \citep{Peeples,Kulkarni_2015}. 

Independent of these limitations, the efficiency of our semi numerical technique shows promising prospects for parameter estimation during helium reionization, especially by using the effective optical depth measurements along different quasar sightlines as the observable. Thus, in a subsequent work we intend to extend our semi-numerical framework to model the HeII Ly$\alpha$ forest effective optical depth ($\tau_{eff}$) from our large grid cells \citep[e.g.][]{Tirth1}. This will allow us to perform parameter estimation on the IGM and source properties using empirical measurements of $\tau_{eff}$ during helium reionization \citep{Worseck19}. The flux power spectrum of the HI Ly$\alpha$ forest has been measured to a high precision by the extended Baryon Oscillation Sky Survey (eBOSS) \citep{eBOSS} and the early data release from the Dark Energy Spectroscopic Instrument (DESI) \citep{DESI1,DESI2}.  
The flux power spectrum is used for putting constraints on the cosmological parameter space at $z \gtrsim 2$ \citep[e.g.][]{cosmo_priya}. Since the photoheating during helium reionization also impacts the HI Ly$\alpha$ forest (and therefore its flux power spectrum), we shall explore the possibility of extending our framework to obtain the HI Ly$\alpha$ forest flux power spectrum in our low resolution simulation. This will allow us to study the impact of incorporating patchy helium reionization on the inferred cosmological parameter constraints. Additionally, our current framework could be extended to other lesser explored observables during helium reionization, for example the hyperfine transition of $^3$HeII at 8.66 GHz \citep{bagla2009hyperfine, Takeuchi, khullar}. Even though the primordial abundance of $^3$He is very less (i.e. $\sim 10^{-5}$ of hydrogen) the large spontaneous rate of decay from its excited state boosts the strength of this signal, which if detected in future observations could serve as a promising direct probe of helium reionization \citep{trott2023first}.

\section*{Acknowledgements}

The authors acknowledge support of the Department of Atomic Energy, Government of India, under project no. 12-R\&D-TFR- 5.02-0700. AK would like to thank Barun Maity for useful discussions. The research made use of the \texttt{CosmoloPy} package \footnote{\url{http://roban.github.com/CosmoloPy/}}. All plots in the paper made use of the \texttt{Matplotlib} package \citep{hunter2007matplotlib}.
\section*{Data Availability}

The data presented in this paper will be shared on reasonable request to the
corresponding author (AK).



\bibliographystyle{JHEP}
\bibliography{Helium} 


\appendix

\section{Luminosity function}
\label{A1} 
The shape of the luminosity function is usually described by a double power law. For a rest frame $1450$ \AA ~ quasar continuum it is given by the following equation at a given redshift: 
\beq
\begin{aligned}
&\Phi(\mr{\mr{M_{1450}}},\mr{z}) = \\ 
& \cfrac{\Phi^*(z)}{10^{0.4(\alpha+1)(\mr{M_{1450}}-M^*_{1450})}+10^{0.4(\beta+1)(\mr{M_{1450}}-M^*_{1450})}},
\end{aligned}
\eeq

where $\Phi^*(z)$ is the amplitude, $M^*_{1450}$ is the break magnitude, $\alpha$ is the bright end slope and $\beta$ is the faint end slope. We use the following form of the redshift evolution of these parameters described by Model 2 of \citep{Girish_UV}:

\beq
\begin{aligned}
log_{10} \phi_* (z) &= F_0( \{ c_{0,j} \},z) \\
M_*(z) &=  F_1( \{ c_{1,j} \},z) \\
\alpha(z) &= F_2( \{ c_{2,j} \},z) \\
\beta(z) &= F_3( \{ c_{3,j} \},z) , \\
\end{aligned}
\eeq
where $F_0$, $F_1$ and $F_2$ are Chebyshev polynomials in $(1+z)$ : 
\beq
F_i(1+z)= \sum_{j=0}^{n_i}c_{i,j} T_j(1+z) ,
\eeq
where $T_j(1+z)$ are the Chebyshev polynomials of the first kind. The faint end slope $\beta$ is described by a double power law:
\beq
F_3(1+z)= c_{3,0} + \cfrac{c_{3,1}}{10^{c_{3,3} \zeta}},
\eeq
where,
\beq
\zeta=log_{10}  \left(\cfrac{1+z}{1+c_{3,1}}\right)
\eeq

 \section{Derivation of the photoheating term}
\label{B}

We provide the full derivation of equation~\ref{temp_ev} here. 
 
The photoheating rate per comoving volume for HeII is given by:
\beq
\epsilon=(1+z)^3 n_{HeII} \int_{\nu_{HeII}}^{\infty} 4 \pi J_{\nu} \sigma_{HeII} (h_P \nu- h_P \nu_{HeII}) \cfrac{d \nu}{h_P \nu},
\label{hui}
\eeq
where, $J_{\nu}$ is the intrinsic flux of photon emission from all quasars, $\sigma_{HeII}$ is the reionization cross-section for HeII ionization, $\nu_{HeIII}$ is the frequency corresponding to the ionization energy of HeII, $h_P$ is the Planck's constant and the comoving number density $n_{HeIII}$ of doubly ionized helium requires multiplication by $(1+z)^3$, since $J_{\nu}$ is evaluated in proper units while our number densities are in units of per comoving volume and need to be converted to proper units.

The rate of change of temperature in IGM if  this energy is distributed over all baryons in the IGM is given by
\beq
\cfrac{dT}{dt} =\cfrac{2}{3 \ k_B} \cfrac{\epsilon}{n_{tot}}
\label{heat}
\eeq

The mean excess energy per HeII ionization is:
\beq
E_J=\cfrac{\int_{\nu_{HeII}}^{\infty} 4 \pi J_{\nu} \sigma_{HeII} (h_P \nu- h_P \nu_{HeII}) d \nu/h_P \nu}{\Gamma_{HeII}},
\eeq
where $\Gamma_{HeII}$ is the photoionization rate for HeII:
\beq
\Gamma_{HeII}=\int_{\nu_{HeII}}^{\infty} 4\pi J_{\nu} \sigma_{HeII,\nu} \cfrac{d \nu}{h_P \nu}
\label{PI}
\eeq
Therefore, the photoheating rate per comoving volume becomes
\beq
\epsilon=(1+z)^3 n_{HeII} \Gamma_{HeII} E_J
\label{ep_ej}
\eeq

In a regime where the radiation is turned on for a timescale which is short (i.e. $t_q$ in our case) compared to the Hubble timescale and the timescale over which recombinations become important:
\beq
\Gamma_{HeII} = \cfrac{1}{\mathrm{x_{HeII}}} \cfrac{d \mathrm{x_{HeIII}}}{dt}.
\label{Gamma}
\eeq

Using the above two equations, the rate of change of temperature due to photoheating:

\beq
\cfrac{dT}{dt}=\cfrac{2}{3 \ k_B} \cfrac{\epsilon}{n_{tot}}=\cfrac{2}{3 \ k_B} \cfrac{n_{He}}{n_{tot}} \cfrac{d \mathrm{x}_{HeIII}}{dt} E_J
\label{neutral}
\eeq

If a gas parcel having an ionization fraction, $\mathrm{x}_{HeIII}$ gets ionized suddenly 
and the excess energy per HeII photoionization is assumed to be shared equally among all baryons, then the rise in temperature (for $ \Delta \mr{x_{HeIII}=1-x_{HeIII}}$), under such a scenario follows from the above equation \citep{Furlanetto_2008,Bolton2009}:
\beq
\Delta T \simeq \cfrac{2}{3 \ k_B} \ 0.035 \ \mathrm{x}_{HeII} E_J
\label{Tinc}
\eeq
Thus, the rise in temperature is determined by the initial neutral fraction that such a gas parcel had before becoming fully ionized.
The factor of 0.035 arises since after the sudden ionization episode, all of helium and hydrogen would be completely ionized due to which the value of $n_{tot}$:
\beq
\begin{aligned}
 n_{tot}&=n_e+n_{HeIII}+n_{HII}
 & =  \cfrac{n_{He}}{Y_P}\left ( 8 - 5 Y_P \right ) ,   
\end{aligned}
\label{ntot}
\eeq
where we used $n_{H}=4 \ n_{He} \ (1-Y_P)/Y_P$.

We define the rise in temperature in equation~\ref{Tinc} if a neutral gas parcel (i.e. $\mathrm{x_{HeII}}=1.0$) gets suddenly reionized: 
\beq
T^{re}_{He} \simeq \cfrac{2}{3 \ k_B} \ 0.035 \ E_J
\label{sudden}
\eeq

Thus, equation~\ref{Tinc} can be re-written as:
\beq
\Delta T= \mathrm{x}_{HeII} T^{re}_{He}
\label{T_red}
\eeq
A lower initial neutral fraction would lead to a lower rise in temperature.

Using the definition of $T^{re}_{He}$ from equation~\ref{sudden} into equation~\ref{neutral} the rate of change of temperature due to a corresponding rate of change of the ionized fraction:
\beq
\cfrac{dT}{dt}=T^{He}_{re} \ \cfrac{d \mathrm{x_{HeIII}}}{dt} 
\label{T_ion_inc}
\eeq
In our large grid cells the change in ionization can be interpreted as a fraction of neutral regions getting ionized in a short interval. Therefore, the above equation describes the corresponding rate of change of temperature in such regions.

The ionization fraction existing in the grid cells are regions which are ionized and are assumed to be in photoionization equilibrium between time-steps, hence the photoionization rate in such regions (where $\mathrm{x}_{HeIII}=1$): 
\beq
\Gamma_{HeII}   = \cfrac{C_{HeIII,i}}{n_{HeII,i}} \ \alpha_B^{HeIII} n_{He,i} {n_{e,i}} (1+z)^3
\label{equilib}
\eeq
Therefore, for ionized regions substituting $\Gamma_{HeII}$ from above into equation~ \ref{ep_ej} and using the definition of $T^{re}_{He}$ from equation~\ref{sudden}:
\beq
\epsilon_i=\cfrac{3 \ k_B}{2 \times 0.035}  T^{He}_{re}  C_{HeIII,i} \alpha_B^{HeIII} {n_{e,i}} {n_{He,i}} (1+z)^6
\eeq

Thus in our grid cells, the photoheating in ionized portions of the cell is $\mathrm{x}_{HeIII,i} \epsilon_i$ which leads to a rate of change of temperature from equation~\ref{heat}:
\beq
\cfrac{dT}{dt}=T^{He}_{re} C_{HeIII,i} \alpha_B^{HeIII}\mathrm{x}_{HeIII,i} \ 2 \ n_{He,i} \left( \cfrac{2}{Y_P} - 1 \right) (1+z)^3,
\eeq
where we used the fact that in regions where hydrogen and helium are fully ionized, $n_{e,i}=\ 2 \ n_{He,i} \left( \cfrac{2}{Y_P} - 1 \right)$.

We are assuming that when helium reionization begins, hydrogen is fully ionized and helium is singly ionized. This implies that hydrogen (and singly ionized helium) is in photo ionization equilibrium with the background photon field. Therefore there will be an existing temperature in such regions due to this ionization equilibrium (equation~A5 of \cite{Maity2022a}): 

\beq
\epsilon_i=3k_B T^{H}_{re} C_{HII,i} \alpha^{HII}_A \chi n_{H,i}^2 (1+z)^6
\label{hyd_eq}
\eeq
The value of $\chi =\cfrac{4-3 Y_P}{4(1-Y_P)}$ in neutral (i.e. regions where helium is singly ionized) and $\chi=\cfrac{2-Y_P}{2(1-Y_P)}$ in ionized regions (i.e. regions where helium is doubly ionized). Therefore from equation~\ref{hyd_eq}, due to ionized hydrogen (and singly ionized helium) in photoionization equilibrium equation~\ref{heat} gives:
\beq
\cfrac{dT}{dt}=T^{H}_{re} C_{HII,i} \alpha_A^{HII} n_{H,i} (1+z)^3,
\eeq
where we used $n_{tot,i}=2 \chi n_{H,i}$.
Thus, the total rate of change of temperature due to photoheating, from equation~\ref{heat}, \ref{neutral} and \ref{hyd_eq} :
\beq
\begin{aligned}
\cfrac{d T_i}{dt} & = T^{He}_{re} C_{HeIII,i} \alpha_B^{HeIII} \mathrm{x}_{HeIII,i} \ 2 \ n_{He,i} \left( \cfrac{2}{Y_P} - 1 \right) (1+z)^3 \\ & + T^{He}_{re}\cfrac{d \mathrm{x_{HeIII}}}{dt}+T^{H}_{re} C_{HII,i} \alpha_A^{HII} n_{H,i} (1+z)^3 
 \label{temp_ev_ap}
\end{aligned}
\eeq

\section{Fitting the Median}
\label{App:fit}
\begin{figure}  
    \centering
     \includegraphics[height=7.9cm,width=8.cm]{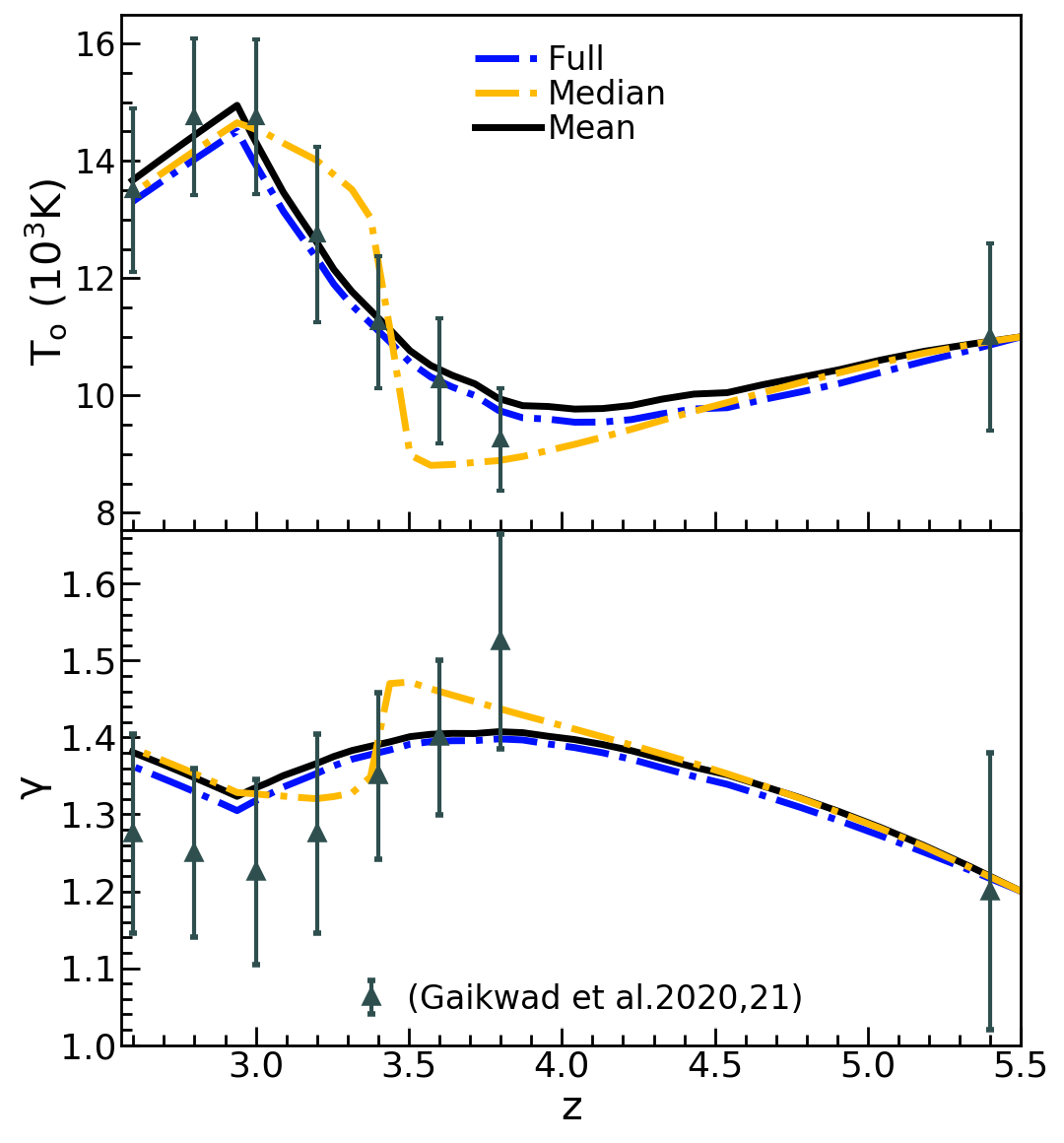} 
     \caption{The plot shows the result of fitting the temperature density power law to the mean (black dashed) and median (yellow) temperature ($log \ T$) in each $log \ \Delta$ bin. We also show the result of directly fitting the temperature density scatter plot in Figure~\ref{sub_scatter}. The grey errorbars are the measured values from \citep{Prakash2020,Prakash21}.}
     \label{fitting}
\end{figure}
In Figure~\ref{fitting} we show the result of fitting the median instead of the mean temperature in the $log~\Delta$ bins. The evolution of the mean is smoother than that of the median, except at the redshift where $T_0$ peaks.  A similar shape for median was also found in the model of \cite{Furlanetto_2008}. The jump in the median values occurs when $\mathrm{x}_{HeIII}=0.5$ i.e. when exactly fifty percent of our sub-grid elements are fully ionized. This is a consequence of sudden reionization of the subgrid elements which leads to two separate branches in temperature for each density bin. On the other hand, the mean can have intermediate temperature value between the hot (ionized) and cold (neutral) sub-grid elements. Once they condense into a single temperature branch after reionization ends, both mean and the median match almost completely. For comparison, we also show the result of fitting the temperature density scatter plot with a power law, without using any binned statistic (blue line). This fit would be largely determined by the region of the scatter plot with the highest density of points (the yellow region in Figure~\ref{sub_scatter}).
 
\section{Convergence analysis} \label{Converge}

 \begin{figure}  
  \centering
   \begin{subfigure}[t]{0.48\textwidth}
   \centering
  \includegraphics[width=\linewidth]{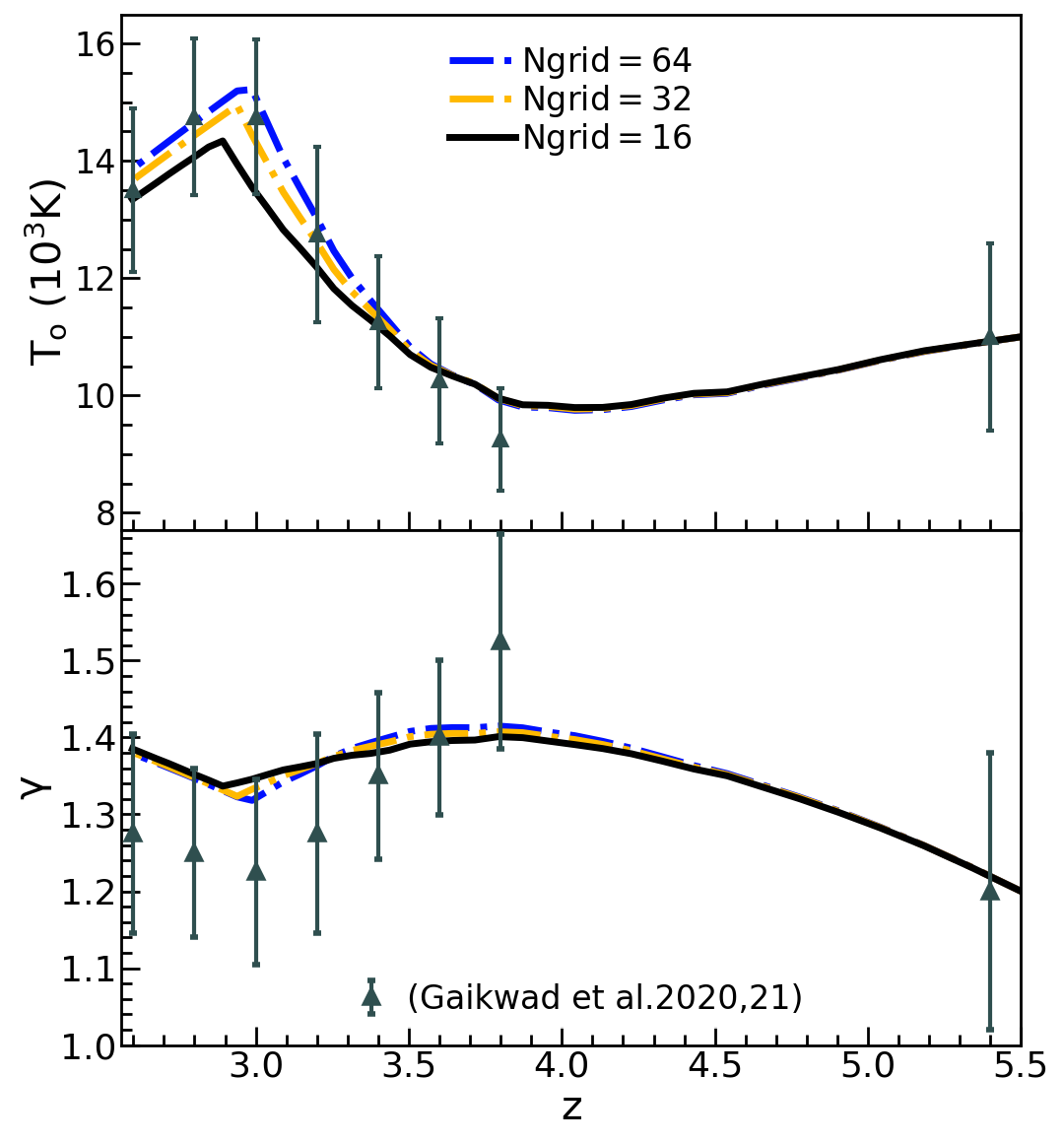}
 \end{subfigure}
   \hfill
   \begin{subfigure}[t]{0.48\textwidth}
     \centering
     \includegraphics[width=\linewidth]{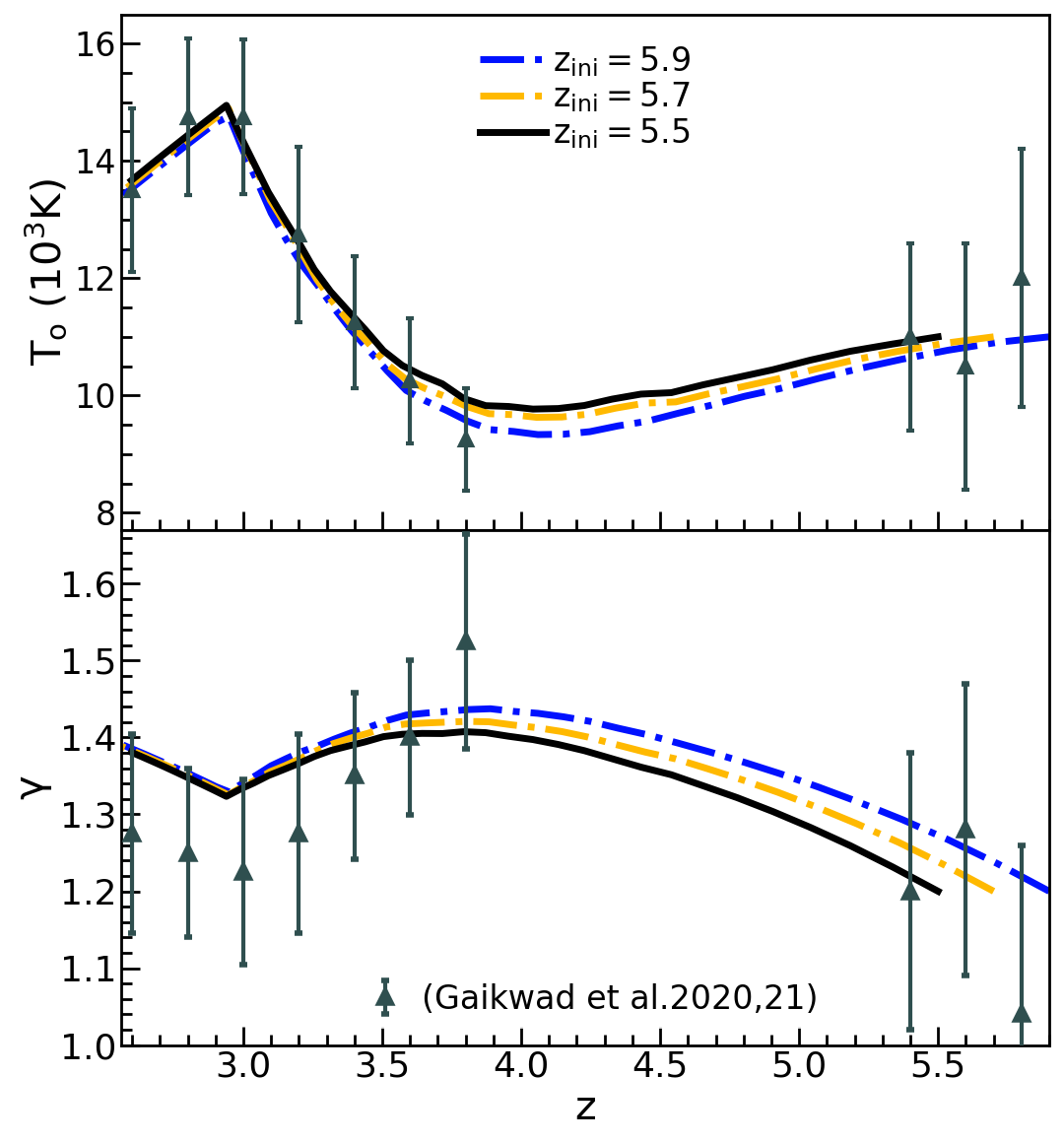}
   
   \end{subfigure}
 \caption{\textit{Left panel:} The plot shows the impact of varying the main grid resolution Ngrid on $T_0$ and $\gamma$ evolution.  \textit{Right panel:} Impact of varying $z_{ini}$, the redshift at which we begin our reionization simulation. The value of $T_0$ and $\gamma$ at $z_{ini}$ is fixed by the measurements of \citep{Prakash2020} which are shown as the grey errorbars at $z=5.4$, $5.6$ and $5.8$ respectively.}
        \label{fig:res_zb}
        \end{figure}
        \begin{figure}  
        \centering
   \begin{subfigure}[t]{0.48\textwidth}
     \centering
     \includegraphics[width=\linewidth]{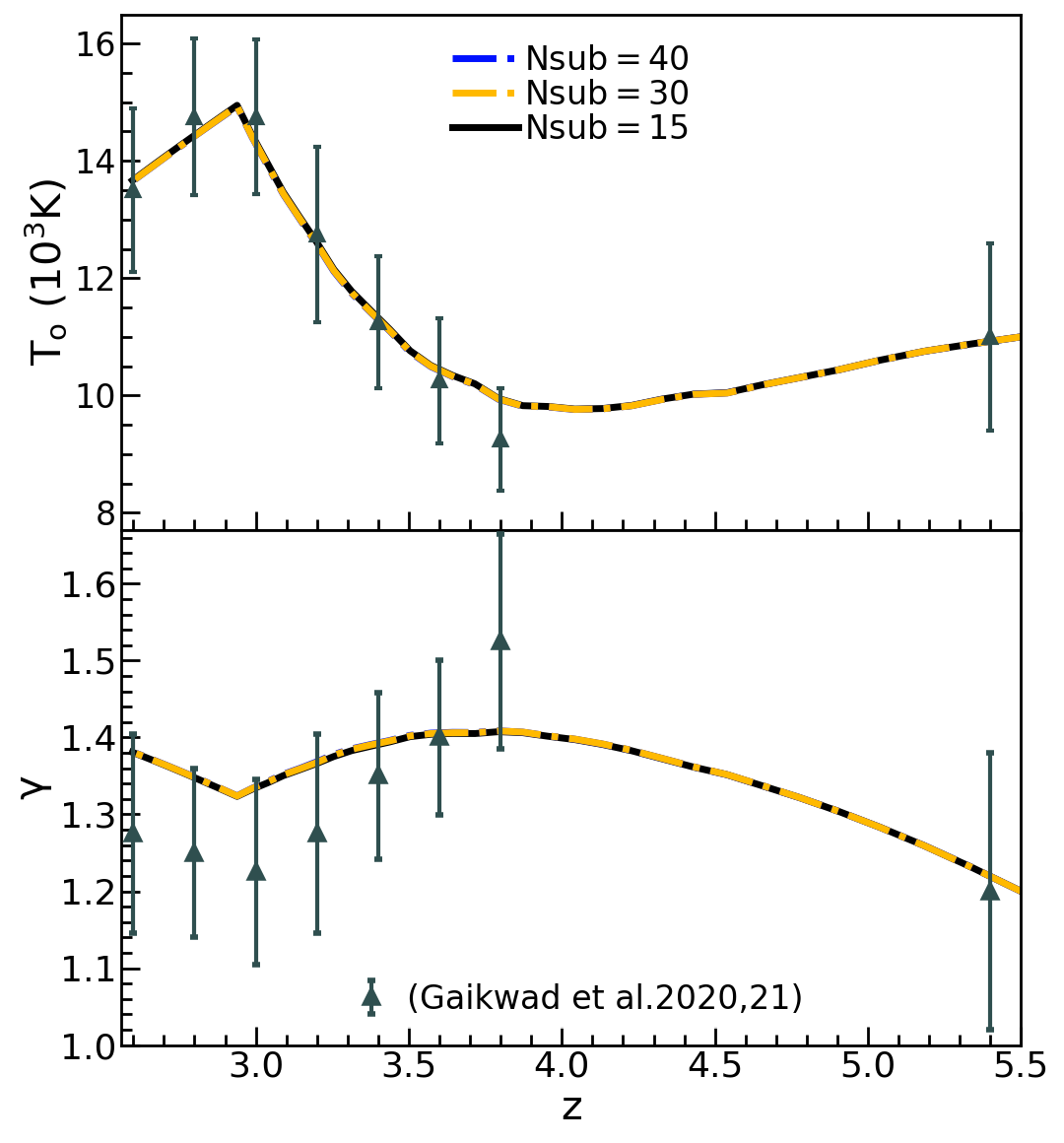}
    
     \label{fig:figure1}
   \end{subfigure}
   \hfill
   \begin{subfigure}[t]{0.48\textwidth}
     \centering
     \includegraphics[width=\linewidth]{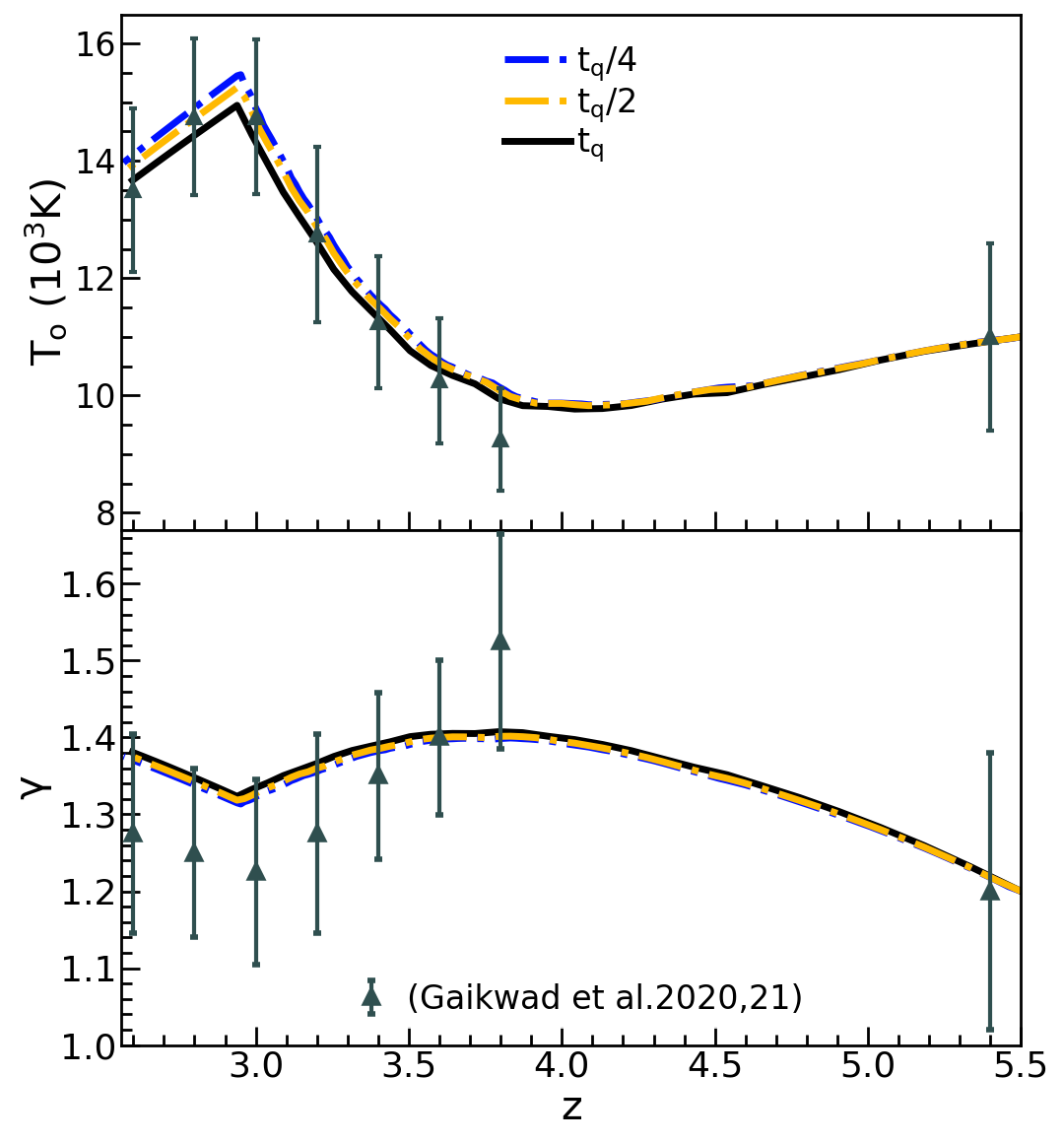}
      \end{subfigure}
   \caption{The plots show the convergence of our subgrid thermal history model, with respect to the number of subgrid elements, $\mathrm{Nsub}$ (\textit{left}) and an increase in time resolution (\textit{Right}). The grey errorbars are the measured values from \citep{Prakash2020,Prakash21}.}
   \label{resolution}
   \end{figure}

In the left panel of Figure~\ref{fig:res_zb} we show the convergence of our results with respect to the number of main grid elements. The dependence on the main grid resolution is a result of our definition of the clumping factor in equation~\ref{eq:clumping}. This is also the reason why we see a mild shift in the position of the peak in the $T_0$ evolution. The values begin to converge beyond $\mathrm{Ngrid=32}$. The right panel of the figure shows the  impact of the redshift $z_{ini}$ at which we start our simulation. The initial temperature density equation of state is set by the respective measurements by \cite{Prakash2020} at $z=5.4, \ 5.6 \ \& \ 5.8$. The magnitude and the redshift at which $T_0$ peaks is independent of the redshift at which we start our simulation for the existing measurements at these starting redshifts. This is because the quasar number densities at these high redshifts is too low and do not vary much over this redshift range to significantly impact helium reionization. This also shows that the thermal memory of hydrogen reionization has almost no impact on the end stages of helium reionization in our simulation.  

The convergence of our subgrid thermal history model with respect to the number of subgrid elements is shown in Figure~\ref{resolution} for our \textit{fiducial} model. We also show convergence with respect to the temporal resolution. For time resolution we divide the simulation time step of $t_q=40$ Myr into $t_q/2=20$ and $t_q/4=10$ Myr (right panel of Figure~\ref{resolution}). We find that our model converges well with respect to both.

\end{document}